\documentclass[aps,twocolumn,pra,superscriptaddress,showpacs,tightenlines]{revtex4}
\usepackage{amssymb}
\usepackage{amsmath}
\usepackage{graphicx}
\usepackage{epsfig}
\usepackage{subfigure}
\usepackage{amsfonts}
\usepackage{CJK}
\usepackage{pgf,tikz}
\usepackage{pgfmath}
\usepackage{psfrag}

\begin{document}

\title{Sudden switch of generalized Lieb-Robinson velocity in a transverse
field Ising spin chain }

\begin{abstract}
The Lieb-Robinson theorem states that the speed at which the correlations
between two distant nodes in a spin network can be built through local
interactions has an upper bound, which is called the Lieb-Robinson
velocity. Our central aim is to demonstrate how to observe the generalized
Lieb-Robinson velocity in an Ising spin chain with a strong transverse
field. Here the generalized Lieb-Robinson velocity is defined as the
correlation propagation speed from a given initial non-correlated
state, which is upper bounded by the Lieb-Robinson velocity. We adopt
and compare four correlation measures for characterizing different
types of correlations, which include correlation function, mutual
information, quantum discord, and entanglement of formation. All the
information-theoretical correlation measures demonstrate the existence
of the generalized Lieb-Robinson velocity. In particular, we find
that there is a sudden switch of the generalized Lieb-Robinson speed
with the increasing of the number of spin.
\end{abstract}

\author{Yu Guo}
\affiliation{Beijing National Laboratory for Condensed Matter Physics, and Institute
of Physics, Chinese Academy of Sciences, Beijing 100190, China}
\affiliation{School of Physics and Electronic Science, Changsha University of
Science and Technology, Changsha 410114, China}
\author{Yang Liu}
\affiliation{Beijing National Laboratory for Condensed Matter Physics, and Institute
of Physics, Chinese Academy of Sciences, Beijing 100190, China}
\author{D. L. Zhou}
\affiliation{Beijing National Laboratory for Condensed Matter Physics, and Institute
of Physics, Chinese Academy of Sciences, Beijing 100190, China}


\maketitle

\section{introduction}
The correlations in a quantum many-body system make its physical properties
can not be regarded as the simple sum of the properties of the composite
subsystems. Recent studies of correlations in quantum information
science show that correlations can be used as indispensable resources
in completing some computing and information tasks \cite{HHHH81}.
Thus the correlations play a key role both in quantum many-body physics
and in quantum information science.

The correlations in a many-body system can be classified into different
types according to different standards, where the two-party correlation
is relatively well understood and widely used in practical problems.
The two-party correlations in a many-body quantum state are measured
by correlation functions associated with physical observables in traditional
physics, while they are measured by the mutual entropy in quantum
information. The two-party correlations can be further classified
into quantum two-party correlations and classical two-party correlations,
and entanglement is a specific type of quantum correlations, which
is extensively investigated in quantum information science. The measures
of different types of corrrelations are proposed in literature. The
widely used entanglement measure is the entanglement of formation
(EoF) \cite{HW97,Wooters98}, and quantum discord (QD) \cite{OZ01,HV01,DSC08}
is used to characterize the quantum correlations.

Recently, many investigations have been made in understanding the
dynamical creation and evolution of correlations between the nearest-neighbor
particle pairs and between two distant particles which are not connected
by direct interactions in the spin chain model, for example, $XX$,
$XY$, and Ising model systems \cite{Cirac08,Sokolovsky08,Sen09,NWu10,Kais11}.
As we know, if two particles directly interact with each other, then
the correlations between these two particles can be built dynamically
from an initial state without correlations. However, if two distant
particles in a quantum network indirectly interact through local interactions,
how fast will the correlations between these two distant particles
be dynamically generated? The Lieb-Robinson theorem \cite{LR72,MBHastingsB04,MBHastingsL04,BHV97,NS06,NYS124}gives
an intriguing answer to this question: The speed of the correlations
between two distant particles has an upper bound, which is called
the Lieb-Robinson velocity. In other words, the correlations outside
the light cone defined by the Lieb-Robinson velocity can be neglected.
Recently, the Lieb-Robinson theorem has received renewed interest
and has been applied to the condensed matter theory and quantum information
theory \cite{BO99,EO97,MMGC82,Poulin,SH81}. For example, it can be
used to derive a general relation on the two-party correlations in
the many-body ground states.

As an upper bound on the speed of correlation generation, can the
Lieb-Robinson velocity be observed in a concrete quantum network?
As far as we know, there is no related experimental report so far.
In fact, we can obtain the Lieb-Robinson velocity through finding
the maximum generalized Lieb-Robinson velocity in all kinds of conditions.
Here the generalized Lieb-Robinson velocity is defined as the correlation
propagation speed from a given initial non-correlated state, which
is upper bounded by the Lieb-Robinson velocity. In this paper, our
central aim is to demonstrate how to observe the generalized Lieb-Robinson
velocity in an Ising spin chain with a strong transverse field. To
solve this problem, we first need to choose a model whose dynamics
relatively easy to simulate. Then we need the measures to characterize
different types of correlations. Finally, we need to give a criterion
to judge whether the correlations between two distant particles appears.
It should be pointed out that the generalized Lieb-Robinson velocity
we study in this article is not the upper bound of speed but the concrete
propagation velocity of correlation. In this article, we choose the
transverse field Ising chain as the basic model, which is exactly
solvable for eigen problems. We consider the measures of different
types of correlations, including correlation functions, mutual entropy,
EoF, and QD. Thus we can investigate whether the generalized Lieb-Robinson
velocity depends on the correlation measures. The criterion for correlation
appearance is to set the correlation measure to a value numerically
so small such that the correlation appearance time is almost fixed.

In our article, we find that the generalized Lieb-Robinson velocity
can be obtained by analyzing a correlation measure. Almost all types
of correlation measures can demonstrate the generalized Lieb-Robinson
velocity in their dynamics. In particular, we find a sudden switch
of the generalized Lieb-Robinson velocities with the increasing of
the spin number. This paper is organized as follows. In Sec. \ref{sec:2},
we will briefly introduce the physical model and the solution of the
dynamics for the reduced two-particle state. In Sec. \ref{sec:3},
we will study and compare the dynamics of different kinds of correlations,
and show the relation between the evolution of correlations and the
length of the spin chain. In Sec. \ref{sec:4}, we will study the
buildup of correlations. Finally, we will give a brief summary.

\section{Model, approximation and solution\label{sec:2}}

We will consider a transverse field Ising spin chain (TFIC) with free
ends, whose Hamiltonian can be written as
\begin{equation}
H=-J\sum_{i=1}^{N-1}\sigma_{i}^{z}\sigma_{i+1}^{z}-B\sum_{i=1}^{N}\sigma_{i}^{x},\label{eq:Isingmodel}
\end{equation}
where $J$ is the coupling constant, $B$ is the strength of the transverse
filed, $N$ is the total number of spin, and $\sigma_{i}^{\alpha}\left(\alpha=x,y,z\right)$
are the Pauli operators acting on the $i$-th spin of the chain.

\begin{figure}[h]
\includegraphics[bb=181 672 420 720, width=7cm]{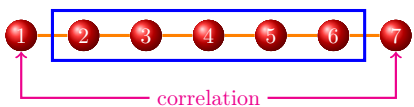}
\caption{Dynamical generation of correlations between the first spin and the
last spin in the spin chain ($N=7$).}\label{fig:1}
\end{figure}

To investigate the dynamical creation of correlations along a TFIC,
we consider the following physical process. First, let the system
stay in the ground state of the system. Then flip the first spin.
Our aim is to observe how the correlations between the first spin
and the last spin will be created dynamically. The physical setting
is demonstrated in Fig. \ref{fig:1}.

\subsection{Rotating-wave approximation for TFIC in a strong magnetic field}

In the weak coupling region ($J\ll B$), the Hamiltonian (\ref{eq:Isingmodel})
can be rephrased in the following more convenient form
\begin{eqnarray}
H&=&-J\sum_{i=1}^{N-1}\left(\sigma_{i}^{-+}\sigma_{i+1}^{+-}+\sigma_{i}^{+-}\sigma_{i+1}^{-+}+\sigma_{i}^{-+}\sigma_{i+1}^{-+}
\right.\notag \\&&\left.
+\sigma_{i}^{+-}\sigma_{i+1}^{+-}\right)-B\sum_{i=1}^{N}\sigma_{i}^{x},
\end{eqnarray}
where $\left|\pm\right\rangle _{i}$ are the two eigenvectors of $\sigma_{i}^{x}$,
and $\sigma_{i}^{\mu\nu}$=$\left|\mu\right\rangle _{i}\left\langle \nu\right|_{i}$
with $\mu,\nu\in\{+,-\}$. Under the rotating-wave approximation (RWA)
\cite{scully}, the Hamiltonian becomes
\begin{equation}
H_{\text{RWA}}=-J\sum_{i=1}^{N-1}\left(\sigma_{i}^{-+}\sigma_{i+1}^{+-}+\sigma_{i}^{+-}\sigma_{i+1}^{-+}\right)-B\sum_{i=1}^{N}\sigma_{i}^{x}.\label{eq:Ising-RWA}
\end{equation}
The Hamiltonian $H_{\text{RWA}}$ is unitarily equivalent to the $XY$ spin model.
Notice that in the above Hamiltonian (\ref{eq:Ising-RWA}), the $x$-component
of the total spin $\sigma_{T}^{x}=\sum_{i=1}^{N}\sigma_{i}^{x}$ is
conserved. In other words, an invariant subspace of the Hamiltonian
can be characterized by a given eigenvalue of $\sigma_{T}^{x}$, and
the dynamics of the system can be studied independently in these invariant
subspaces.

\begin{figure}[h]
\centering
\subfigure{
    \includegraphics[width=6.5cm, clip]{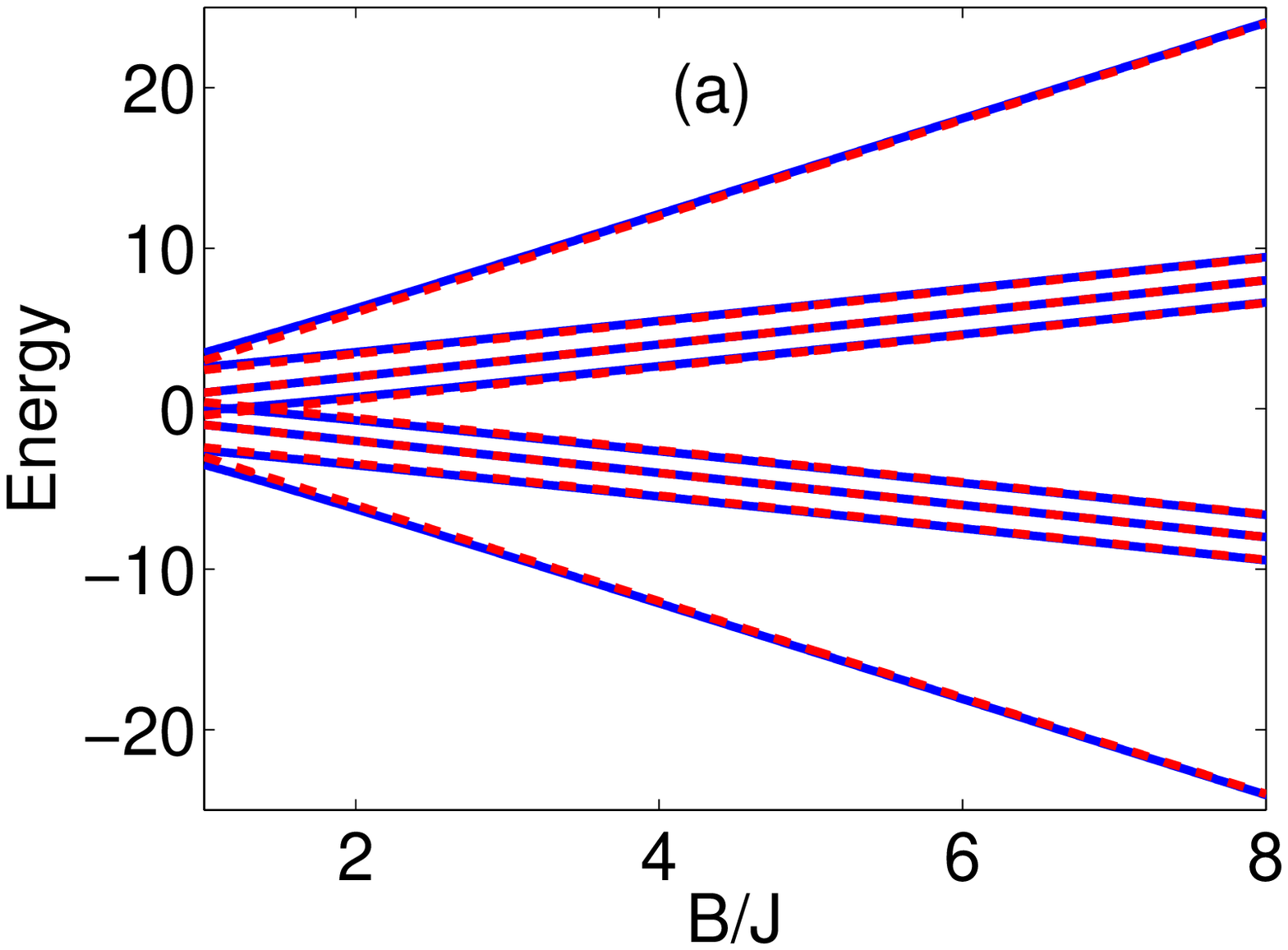}}
\subfigure{
    \includegraphics[width=6.5cm, clip]{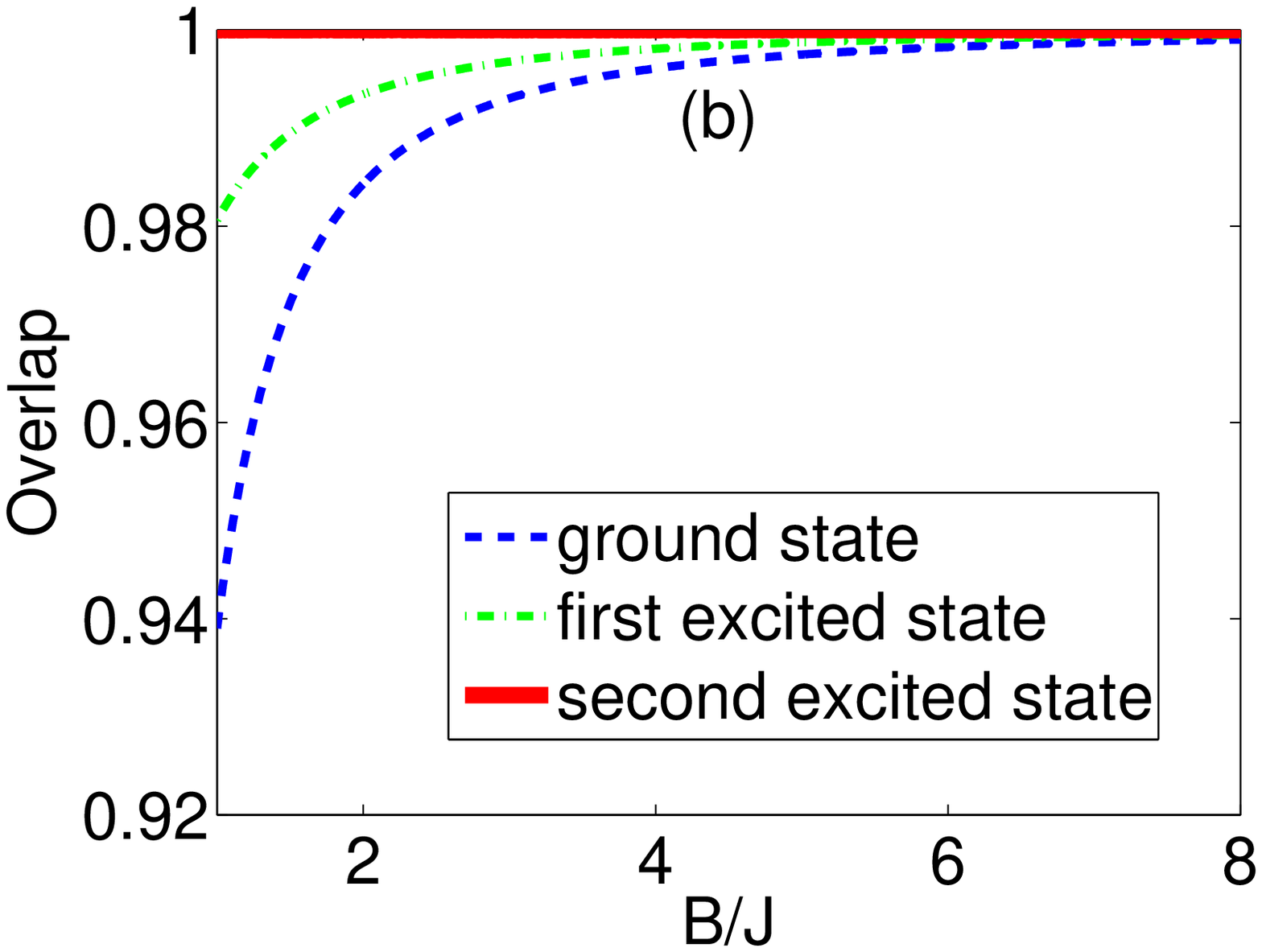}}
\caption{(Color online) (a) The energy spectrum ($N=3$) as a function of $B/J$.
Dashed lines: the result in RWA; Solid lines: the result without RWA.
(b) The overlap between the wave function obtained with RWA and without
RWA as a function of $B/J$. Dashed line: the overlap of ground states;
Dotted-dashed line: the overlap of the first exctied states; Sloid
line: the overlap of the second excited states.}\label{fig:2}
\end{figure}

Before investigating further the dynamics of the system, we directly
check the validity of RWA for the Hamiltonian by numerically comparing
the eigenvalues and the eigenvectors between the Hamiltonian without
RWA (\ref{eq:Isingmodel}) and the one with RWA (\ref{eq:Ising-RWA}).
The numerical results for $N=3$ are demonstrated in Fig. \ref{fig:2}.
As expected, we find that RWA is an excellent approximation as long
as $B\gg J.$

\subsection{Quantum state evolution under RWA}

In this section, we will obtain the analytical results on the quantum
state evolution of the system under RWA.

In the ground state of the Hamiltonian $H_{\text{RWA}}$, all the spins point along
the positive direction of $x$ axis. While flipping the first spin,
we get the initial state $\vert\psi(0)\rangle=\vert-++\cdots+\rangle$.
Because $\sigma_{T}^{x}\vert\psi(0)\rangle=(N-2)\vert\psi(0)\rangle$
and $[\sigma_{T}^{x},H_{\text{RWA}}]=0$, the quantum state will evolve with time
in the eigen space of $\sigma_{T}^{x}$ with eigenvalue $N-2$. All
the eigenvectors of the subspace can be denoted as $\vert m\rangle=\prod_{i=1}^{m-1}\vert+\rangle_{i}\vert-\rangle_{m}\prod_{j=m+1}^{N}\vert+\rangle_{j}$
for $m\in\{1,2,\cdots,N\}$. Using these new notations, the initial
state $\vert\psi(0)\rangle=\vert1\rangle$, and the Hamiltonian in
this subspace
\begin{equation}
H=-(N-2)B-J\left(\begin{array}{ccccc}
0 & 1 & 0 & \cdots & 0\\
1 & 0 & 1 & \cdots & 0\\
0 & 1 & 0 & \cdots & 0\\
\vdots & \vdots & \vdots & \ddots & 1\\
0 & 0 & 0 & 1 & 0
\end{array}\right).
\end{equation}
The Hamiltonian $H_{\text{RWA}}$ has eigenstates
\begin{equation}
\left|\psi_{k}\right\rangle =\sqrt{\frac{2}{N+1}}\sum_{n=1}^{N}\sin\left(\frac{\pi kn}{N+1}\right)\left|n\right\rangle ,
\end{equation}
with eigenvalues
\begin{equation}
E_{k}=-(N-2)B+2J\cos\frac{k\pi}{N+1},
\end{equation}
for $k=1,2,\cdots,N$.

The quantum state of the system at time $t$ is given by
\begin{eqnarray}
\left|\Psi(t)\right\rangle  & = & \sum_{n=1}^{N}A_{n}\left|n\right\rangle ,
\end{eqnarray}
where
\begin{equation}
A_{n}=\frac{2}{N+1}\sum_{k=1}^{N}e^{-iE_{k}t}\sin\left(\frac{\pi k}{N+1}\right)\sin\left(\frac{\pi kn}{N+1}\right).\label{eq:An}
\end{equation}

Since we will consider the correlations between the first spin and
the last spin, we only need the reduced density matrix of these two
spins, which is given by

\begin{equation}
\rho_{1N}=\left(\begin{array}{cccc}
0 & 0 & 0 & 0\\
0 & \left|A_{1}\right|^{2} & A_{1}^{\ast}A_{N} & 0\\
0 & A_{1}A_{N}^{\ast} & \left|A_{N}\right|^{2} & 0\\
0 & 0 & 0 & 1-\left|A_{1}\right|^{2}-\left|A_{N}\right|^{2}
\end{array}\right).\label{eq:rho1n}
\end{equation}
The above formula implies that only $A_{1}$ and $A_{N}$ are needed
to be calculated for our purpose, which reduces a large amount of
computation.

\section{dynamics of different types of correlations\label{sec:3} }

To study the correlations between the first spin and the last spin,
we adopt both traditional method and information method to characterize
the degrees of correlations. In this section, we will numerically
study the dynamical evolution of different measures of correlations
between the first spin and the last spin. The correlation measures
we adopt include the correlation functions, the mutual information,
quantum discord, and the entanglement of formation. We aim to find
out the relation between the evolution of correlations and the length
of spin chain.

\subsection{Traditional method: Correlation function}

Correlation function (CF) is a traditional tool in describing the
correlation effects in a many-body system. For a two-qubit system,
the CFs are defined by \cite{DZ10}
\begin{equation}
C_{F}\left(\sigma_{1}^{\alpha},\sigma_{N}^{\beta}\right)=\mathrm{Tr}\left(\rho_{1N}\sigma_{1}^{\alpha}\sigma_{N}^{\beta}\right)-\mathrm{Tr}\left(\rho_{1}\sigma_{1}^{\alpha}\right)\mathrm{Tr}\left(\rho_{N}\sigma_{N}^{\beta}\right),\label{eq:CF}
\end{equation}
where $\alpha,\beta\in\left\{ x,y,z\right\} $, and $\rho_{1}$ and
$\rho_{N}$ are the reduced density matrices of the bipartite quantum
state $\rho_{1N}.$ Inserting Eq. (\ref{eq:rho1n}) into Eq. (\ref{eq:CF}),
we have
\begin{eqnarray}
C_{F}\left(\sigma_{1}^{z},\sigma_{N}^{z}\right) & = & -4\left|A_{1}\right|^{2}\left|A_{N}\right|^{2},\label{eq:CFzz}\\
C_{F}\left(\sigma_{1}^{x},\sigma_{N}^{x}\right) & = & A_{1}A_{N}^{\ast}+A_{1}^{\ast}A_{N}.\label{eq:CFxx}
\end{eqnarray}

Here we find that $C_{F}\left(\sigma_{1}^{x},\sigma_{N}^{x}\right)=0$
when $N$ is even. This result can be proved as follows. We have the
relation
\begin{widetext}
\begin{eqnarray}
C_{F}\left(\sigma_{1}^{x},\sigma_{N}^{x}\right) & = & A_{1}A_{N}^{\ast}+A_{1}^{\ast}A_{N} \nonumber \\
 & = & \frac{8}{(N+1)^{2}}\sum_{k,m=1}^{N}\cos\left[\left(E_{k}-E_{m}\right)t\right]\left(-1\right)^{m+1}\sin^{2}\left(\frac{k\pi}{N+1}\right)\sin^{2}\left(\frac{m\pi}{N+1}\right) \nonumber\\
 & = & \frac{8}{(N+1)^{2}}\sum_{k,m=1}^{N}\cos\left[2J\left(\cos\frac{m\pi}{N+1}-\cos\frac{k\pi}{N+1}\right)t\right]\nonumber\\
 &  & \times\left(-1\right)^{m+1}\sin^{2}\left(\frac{k\pi}{N+1}\right)\sin^{2}\left(\frac{m\pi}{N+1}\right)\nonumber\\
 & = & \frac{8}{(N+1)^{2}}\sum_{l=1}^{N}\sum_{k=1}^{N}\cos\left[2J\left(\cos\frac{(N+1-l)\pi}{N+1}-\cos\frac{(N+1-k)\pi}{N+1}\right)t\right]\nonumber\\
 &  & \times\left(-1\right)^{(N+1-l)+1}\sin^{2}\left(\frac{(N+1-k)\pi}{N+1}\right)\sin^{2}\left(\frac{(N+1-l)\pi}{N+1}\right)\nonumber\\
 & \text{}= & (-1)^{N+1}C_{F}(\sigma_{1}^{x},\sigma_{N}^{x}).
\end{eqnarray}
\end{widetext}
When $N$ is even, we have $C_{F}(\sigma_{1}^{x},\sigma_{N}^{x})=-C_{F}(\sigma_{1}^{x},\sigma_{N}^{x})$,
which implies that $C_{F}(\sigma_{1}^{x},\sigma_{N}^{x})=0$.

It is worth pointing out that $C_{F}(\sigma_{1}^{z},\sigma_{N}^{z})\neq0$
whether $N$ is odd or even. These different correlation behaviors
between $C_{F}(\sigma_{1}^{x},\sigma_{N}^{x})$ and $C_{F}(\sigma_{1}^{z},\sigma_{N}^{z})$
shows that the correlations measured by correlation functions depend
on which correlation function we choose. In the viewpoint of quantum
information, correlation in a two-partite quantum state is a property
of the state, and the characterization of the correlation does not
depend on what measurements we obtain the correlation. This information
viewpoint will be detailedly discussed in the next section.

\begin{figure}[tbb]
\centering
\subfigure{
    \includegraphics[width=4cm]{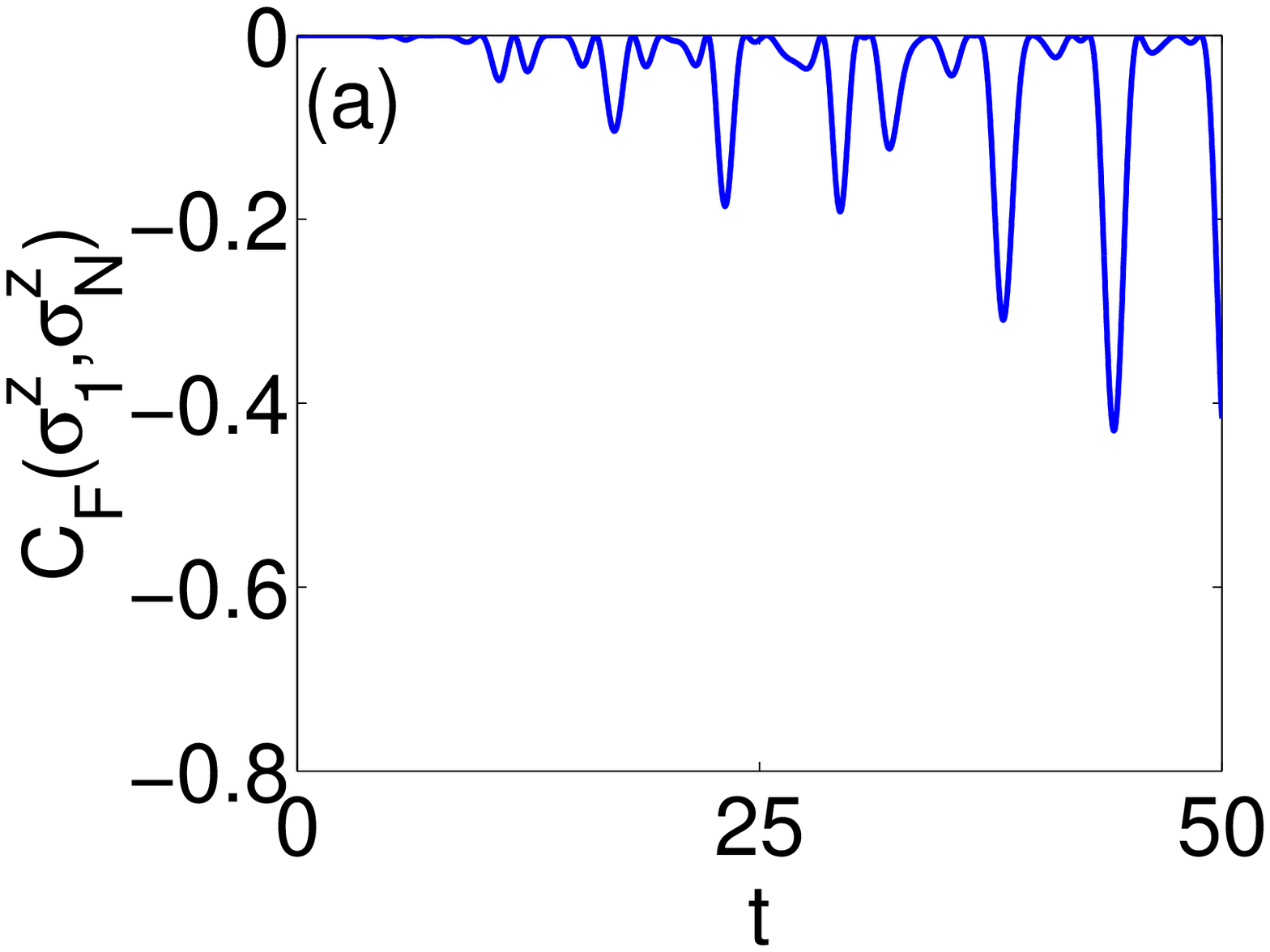}}
\subfigure{
    \includegraphics[width=4cm, clip]{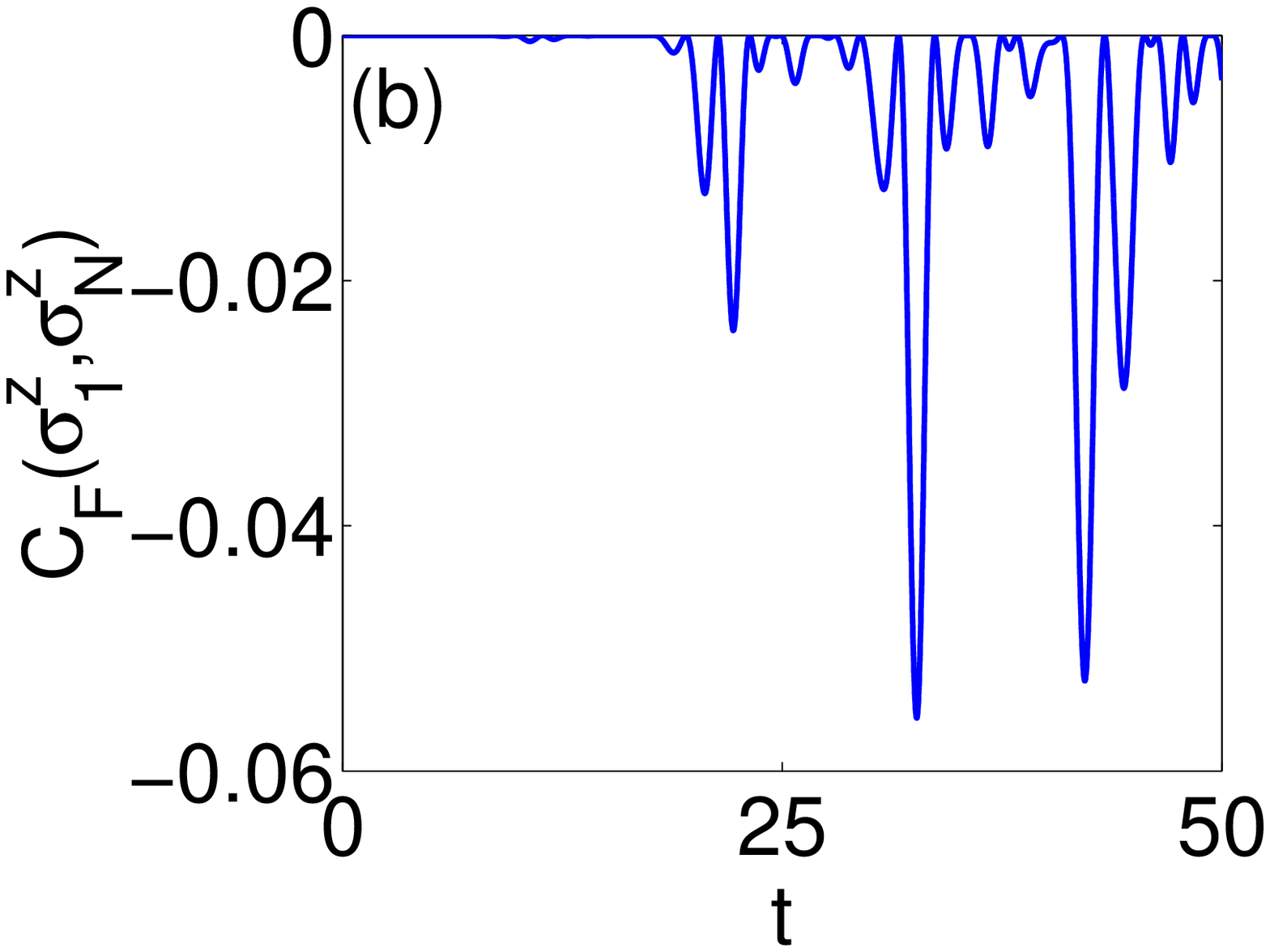}}
\subfigure{
    \includegraphics[width=4cm]{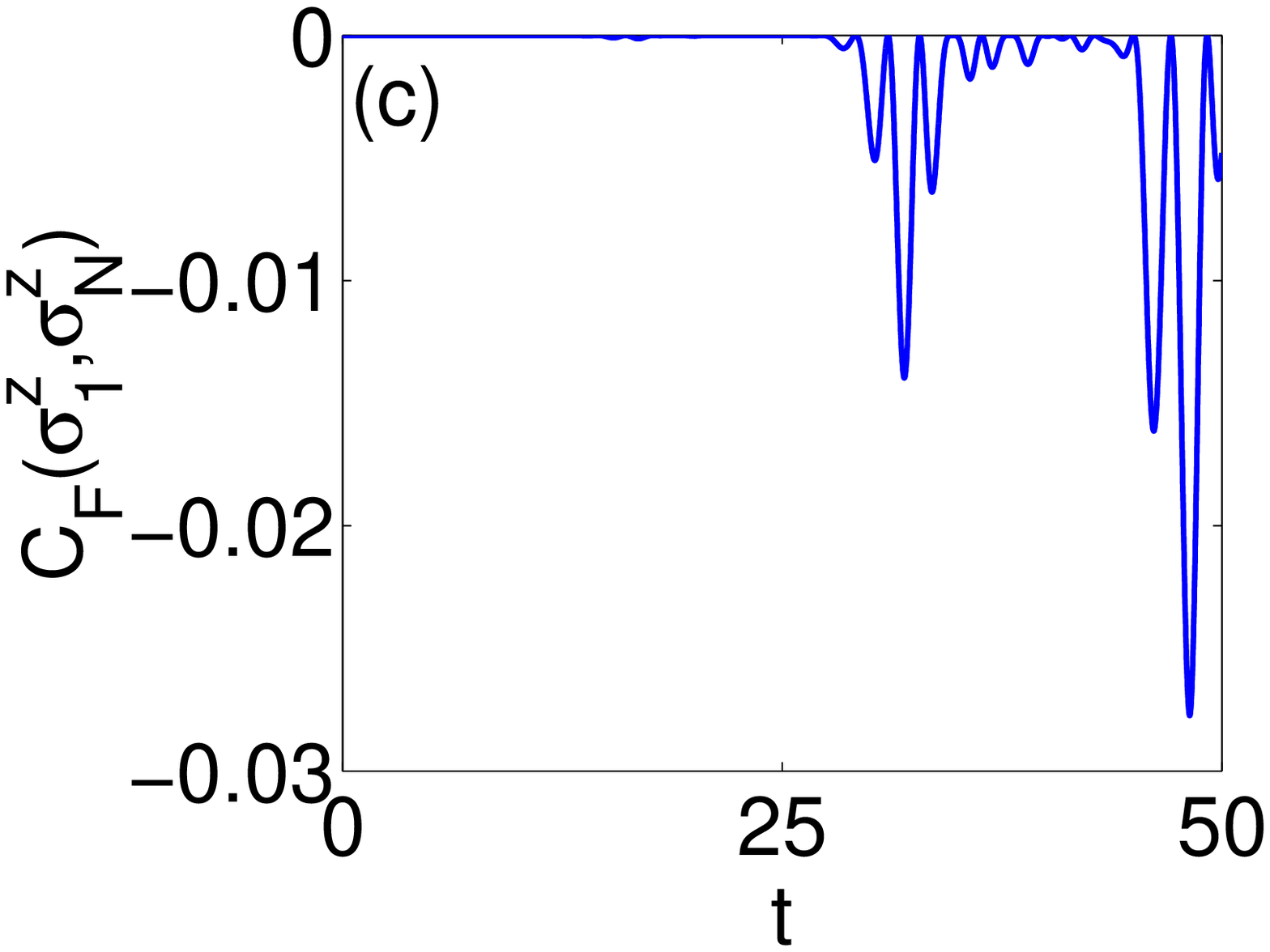}}
\subfigure{
    \includegraphics[width=4cm, clip]{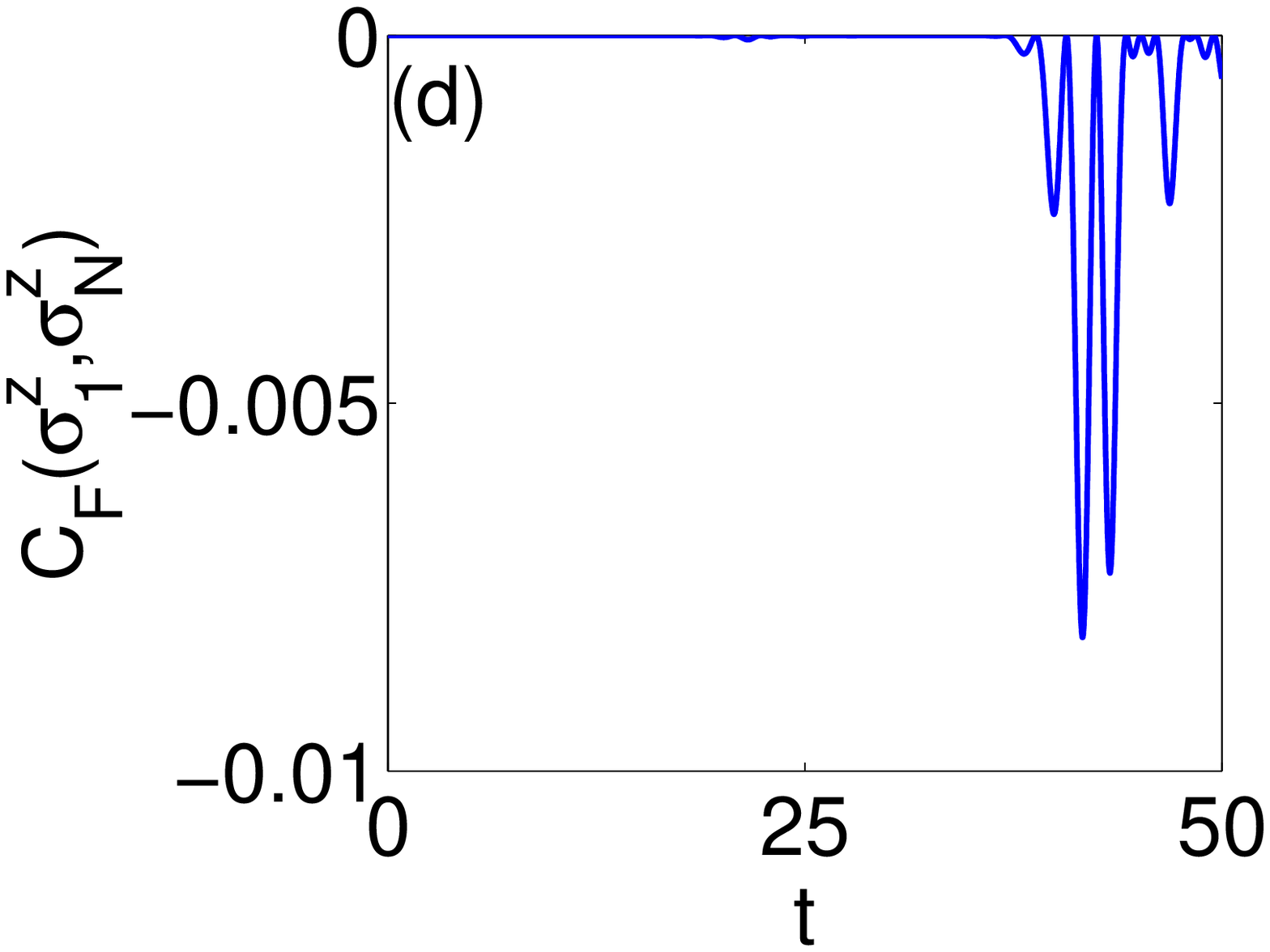}}
\caption{Dynamics of correlation function $C_{F}\left(\sigma_{1}^{z},\sigma_{N}^{z}\right)$
of the spin chain with different lengths: (a) $N=10$, (b) $N=20$,
(c) $N=30$, (d) $N=40$.} \label{fig:3}
\end{figure}
\begin{figure}[tbb]
\centering
\subfigure{
    \includegraphics[width=4cm]{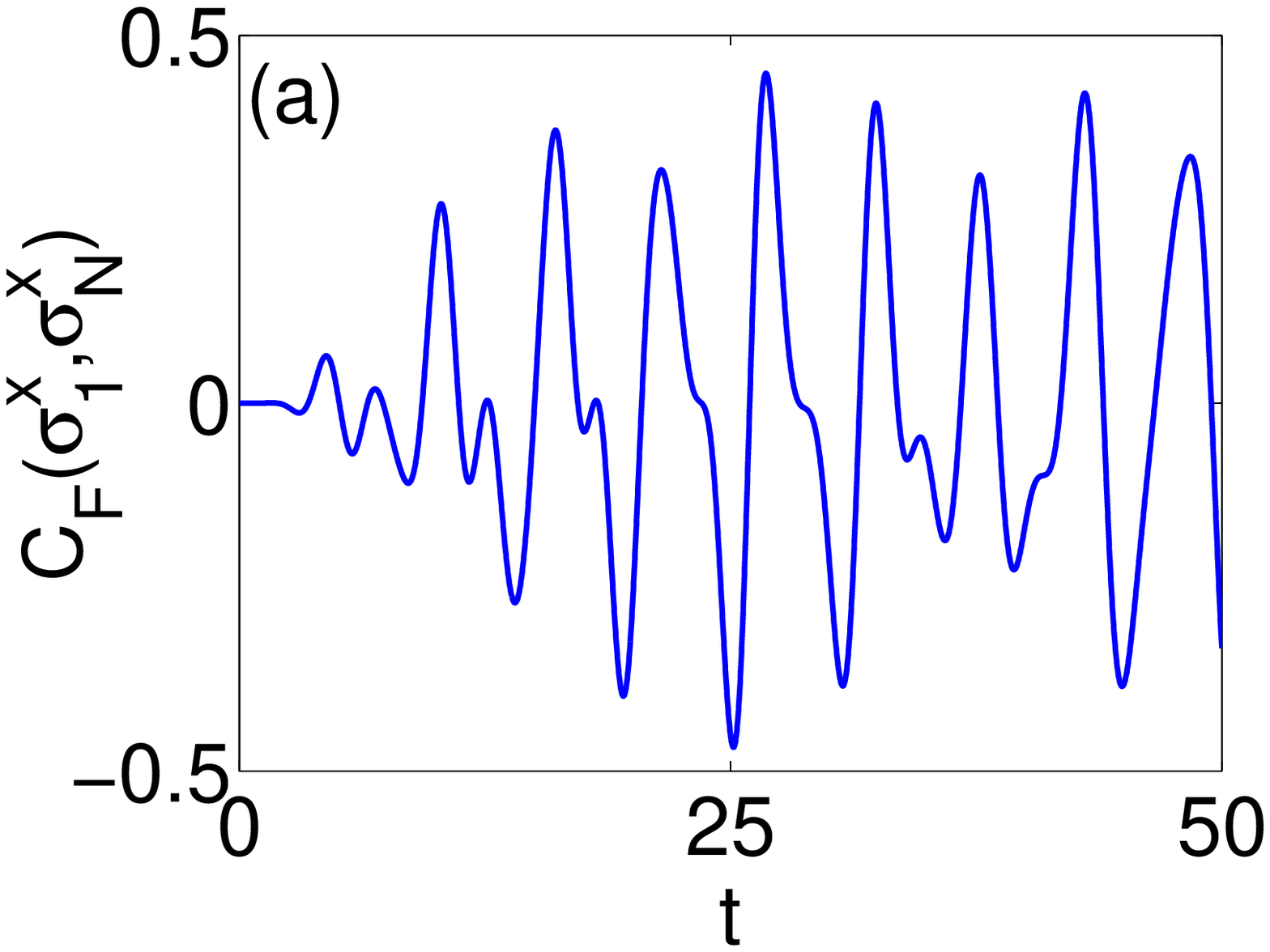}}
\subfigure{
    \includegraphics[width=4cm, clip]{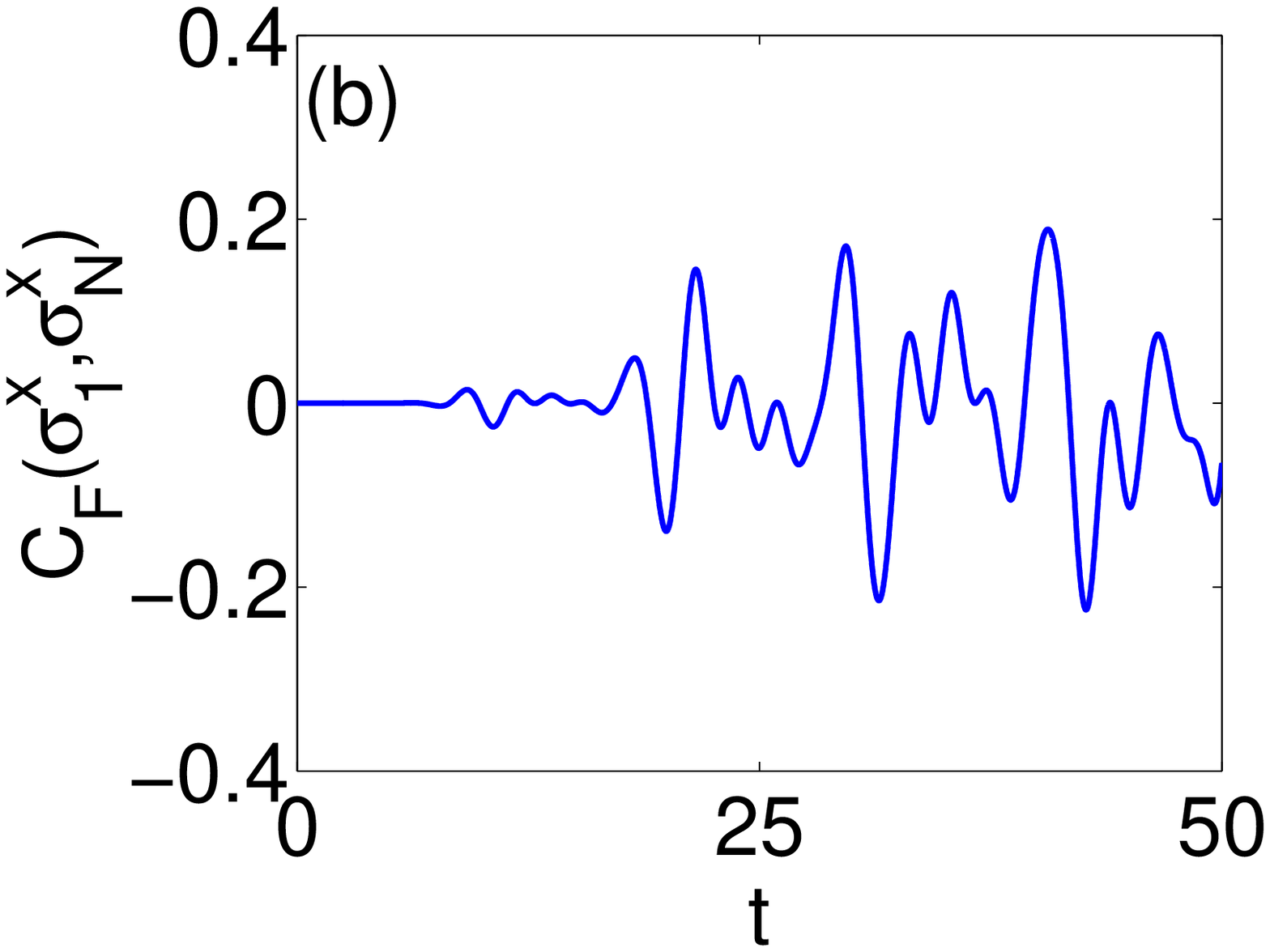}}
\subfigure{
    \includegraphics[width=4cm]{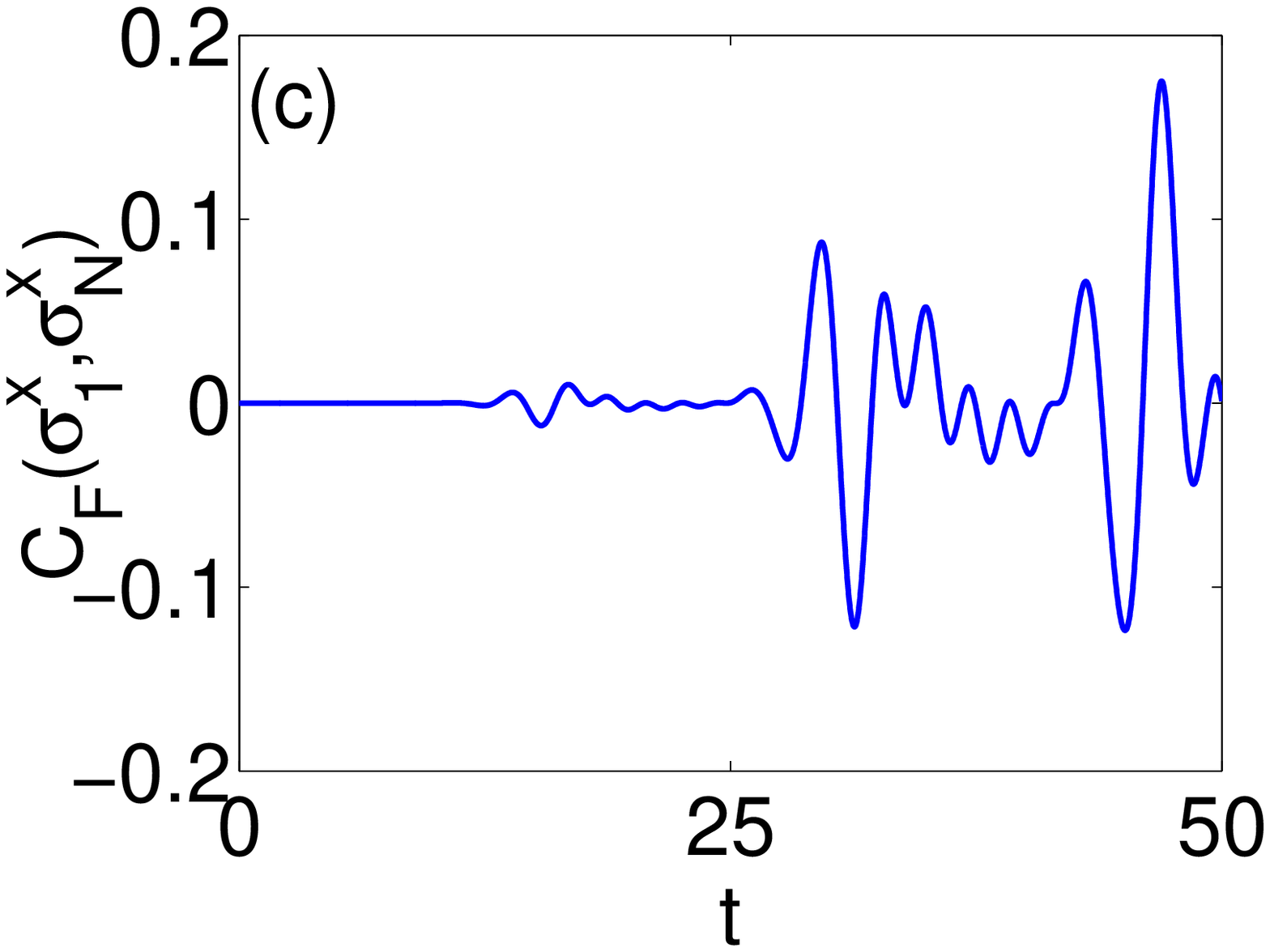}}
\subfigure{
    \includegraphics[width=4cm, clip]{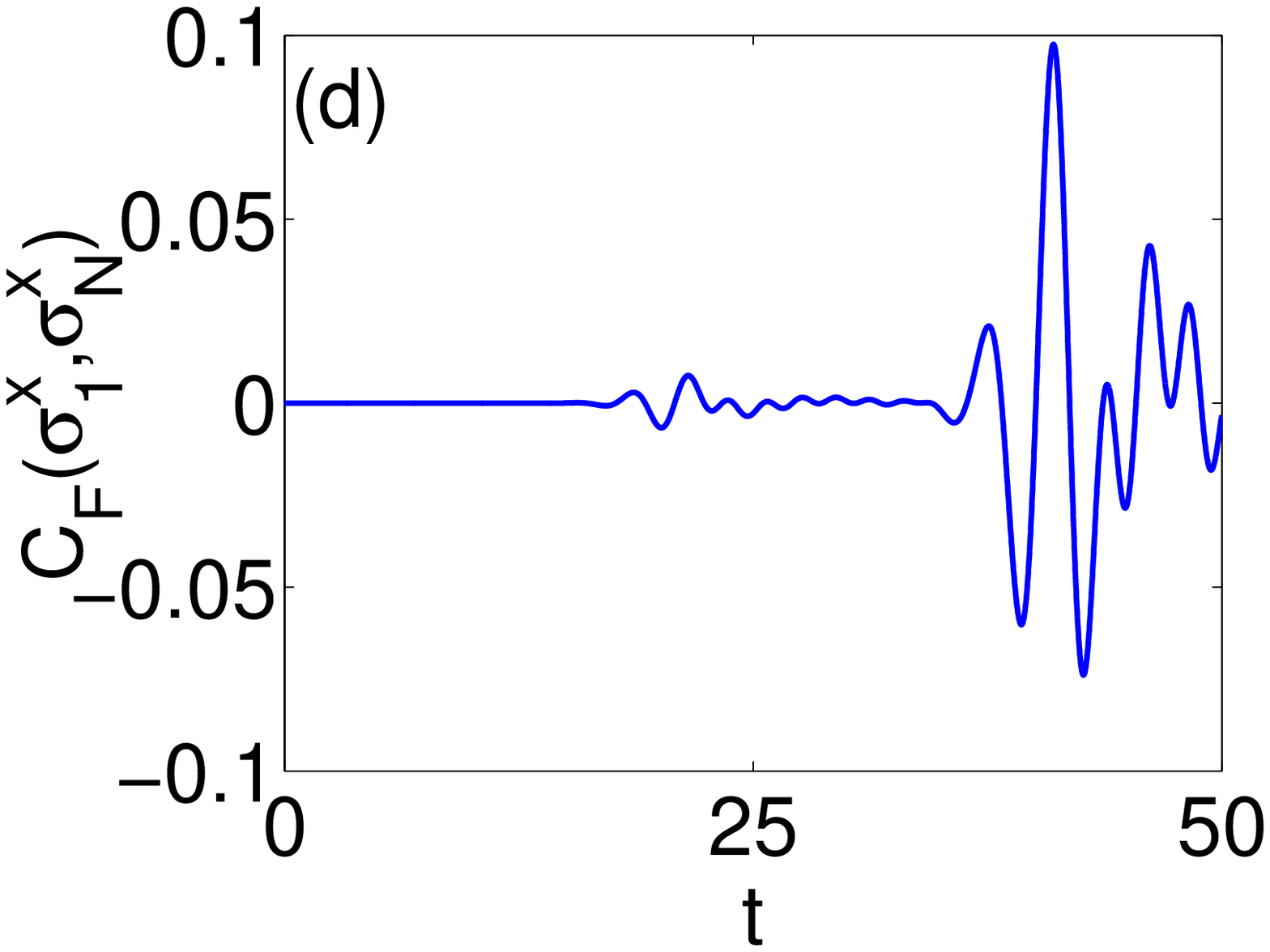}}
\caption{Dynamics of correlation function $C_{F}\left(\sigma_{1}^{x},\sigma_{N}^{x}\right)$
of the spin chain with different lengths: (a) $N=9$, (b) $N=19$,
(c) $N=29$, (d) $N=39$. }\label{fig:4}
\end{figure}
The numerical results about the dynamical evolution of the two CFs,
$C_{F}\left(\sigma_{1}^{z},\sigma_{N}^{z}\right)$ and $C_{F}\left(\sigma_{1}^{x},\sigma_{N}^{x}\right)$,
of the spin chain with different lengths are shown in Fig. \ref{fig:3}
and Fig. \ref{fig:4}. From the two figures, we can observe that the
creation of correlations between the first spin and the last spin,
which have no direct interactions, is not instantaneous but requires
the formation time, and the formation time is directly proportional
to the length of spin chain. In other words, the longer the distance
between the first spin and the last spin is, the more time the appearance
of their correlation needs.

\subsection{Information method: Mutual information, quantum discord, entanglement
of formation}

\begin{figure}[t]
\includegraphics[bb=241 594 379 723, width=4cm]{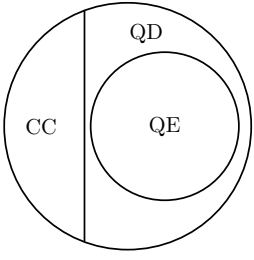}
\caption{Demonstration of classification of the total correlation meausred
by mutual entropy. Classical correlation, quantum discord, quantum
entanglement are abbraviated as CC, QD, and QE respectively.} \label{fig:5}
\end{figure}

In quantum information science, the characterization of correlations
in a bipartite quantum state is relatively well understood. The degree
of the total correlation in a bipartite quantum state is measured
by the mutual entropy. The total correlation can be classified into
classical correlation and quantum discord \cite{OZ01,HV01}. Quantum
entanglement is a special type of quantum discord. The classification
of correlations is demonstrated in Fig. \ref{fig:5}.

The total correlation between two subsystems A and B are quantified
by the mutual information (MI)
\begin{equation}
\mathcal{I}(\rho^{AB})=S(\rho^{A})+S(\rho^{B})-S(\rho^{AB}),
\end{equation}
where $S(\rho)\equiv-\mathrm{Tr}(\rho\log_{2}\rho)$ is the von Neumann
entropy. To define the classical correlation contained in the state
$\rho^{(AB)}$, we consider the following process. B performs a projective
measurement $\{\Pi_{j}\}$ on the subsystem B, and will get a state
\begin{equation}
\rho_{k}^{A}=\frac{1}{p_{k}}\mathrm{Tr_{B}}\left(I\otimes\Pi_{k}\rho^{AB}I\otimes\Pi_{k}\right),
\end{equation}
with the corresponding probability $p_{k}=\mathrm{Tr_{AB}}\left(I\otimes\Pi_{k}\rho^{AB}I\otimes\Pi_{k}\right)$.
By performing a measurement on subsystem A, A wants to give the information
on which measurement result got by B. The upper bound of the information
is the Holevo bound
\begin{equation}
\chi(\rho^{AB}|\{\Pi_{k}\})=S(\rho^{A})-p_{k}S(\rho_{k}^{A}).
\end{equation}
Then a measure of classical correlation (CC) in the state $\rho^{(AB)}$
is defined by
\begin{equation}
\mathcal{J}(\rho^{AB})=\max_{\{\Pi_{k}\}}\chi(\rho^{AB}|\{\Pi_{k}\}).\label{eq:cc}
\end{equation}
Once CC is obtained, the QD is obtained by subtracting CC from the
MI
\begin{equation}
\mathcal{D}(\rho^{AB})=\mathcal{I}(\rho^{AB})-\mathcal{J}(\rho^{AB}).\label{eq:qd}
\end{equation}

Quantum entanglement is a special type of quantum discord, which is
widely investigated in quantum information. One of the most useful
measures of entanglement is entanglement of formation \cite{HW97,Wooters98},
which is defined by
\begin{equation}
\mathcal{E}(\rho^{(AB)})=\frac{1}{2}\min_{\{p_{i},\vert\psi_{i}\rangle\}}p_{i}\mathcal{I}(\vert\psi_{i}^{(AB)}\rangle),
\end{equation}
where $\rho^{(AB)}=\sum_{i}p_{i}\vert\psi_{i}^{(AB)}\rangle\langle\psi_{i}^{(AB)}\vert$.

For a two-qubit system, all the above correlation measures can be
obtained analytically or numerically. For example, in our case, the
mutual information is
\begin{eqnarray}
\mathcal{I}(1,N) & = & \left|A_{N}\right|^{2}\log_{2}\left[\frac{\left(1-\left|A_{N}\right|^{2}\right)\left(\left|A_{1}\right|^{2}+\left|A_{N}\right|^{2}\right)}{\left|A_{N}\right|^{2}\left(1-\left|A_{1}\right|^{2}-\left|A_{N}\right|^{2}\right)}\right]
\nonumber \\ &  &
 +\left|A_{1}\right|^{2}\log_{2}\left[\frac{\left(1-\left|A_{1}\right|^{2}\right)\left(\left|A_{1}\right|^{2}
+\left|A_{N}\right|^{2}\right)}{\left|A_{1}\right|^{2}\left(1-\left|A_{1}\right|^{2}-\left|A_{N}\right|^{2}\right)}\right]
\nonumber \\ &  & +\log_{2}\left[\frac{1-\left|A_{1}\right|^{2}-\left|A_{N}\right|^{2}}{\left(1-\left|A_{1}\right|^{2}\right)\left(1-\left|A_{N}\right|^{2}\right)}\right],
\end{eqnarray}
which is shown in Fig. 6. Following the method introduced in Ref.
\cite{ali,luo1}, we can numerically obtain the dynamical evolution
of QD as shown in Fig. 7. In order to compare the dynamical evolution
between MI(QD) and EoF, we also plot the dynamical evolution of EoF
as shown in Fig. 8.

\begin{figure}[tbh]
\centering
\subfigure{
    \includegraphics[width=4cm]{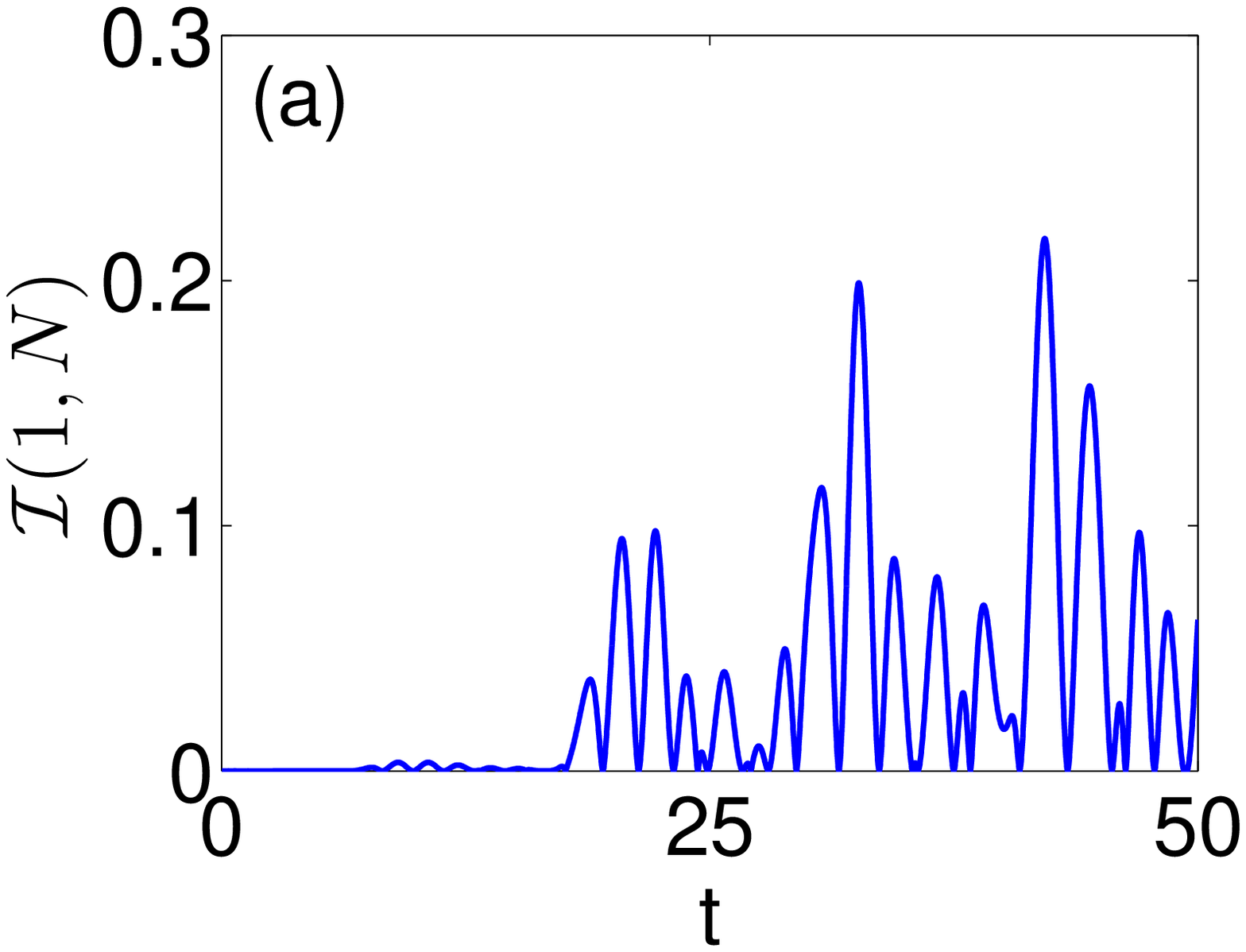}}
\subfigure{
    \includegraphics[width=4cm, clip]{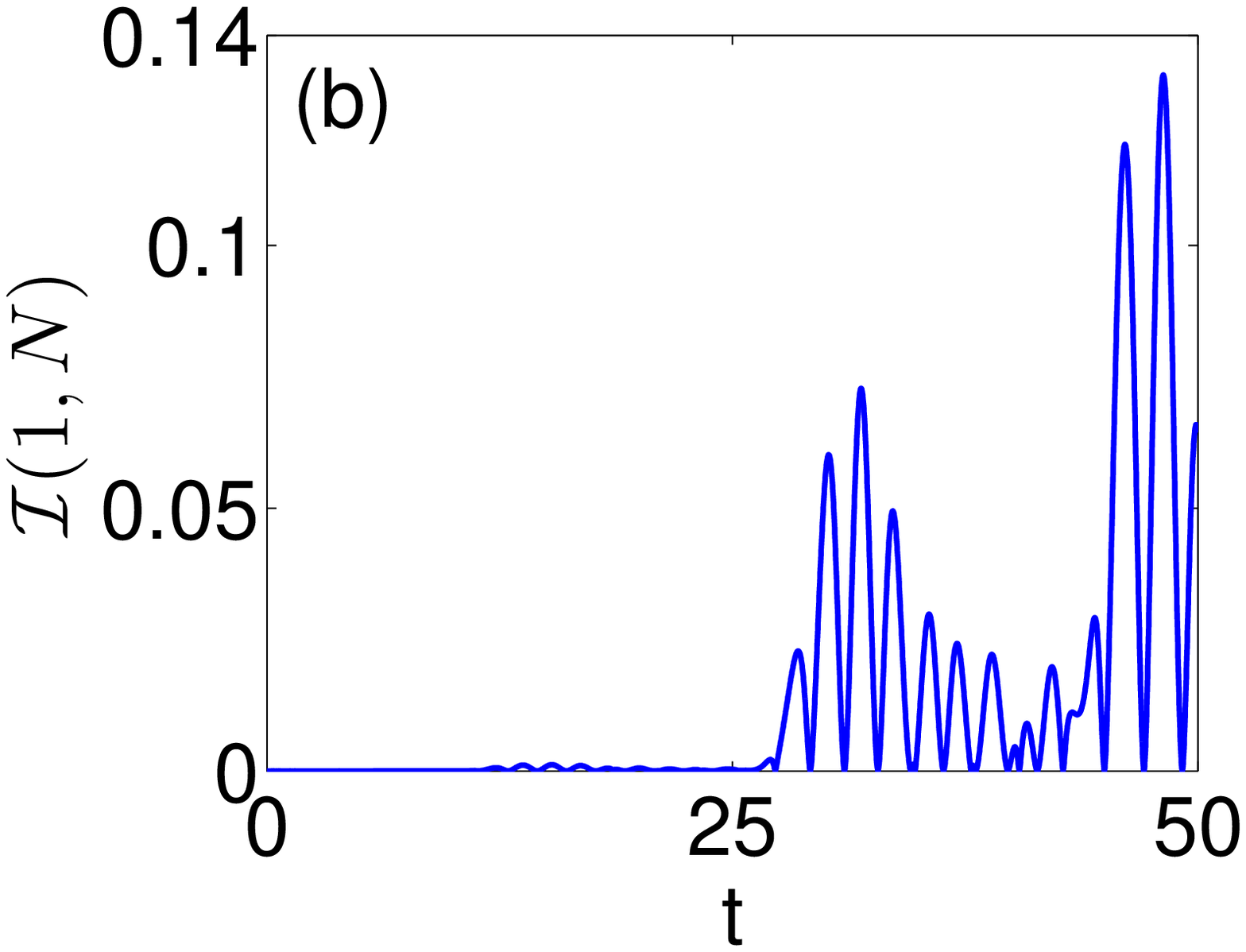}}
\subfigure{
    \includegraphics[width=4cm]{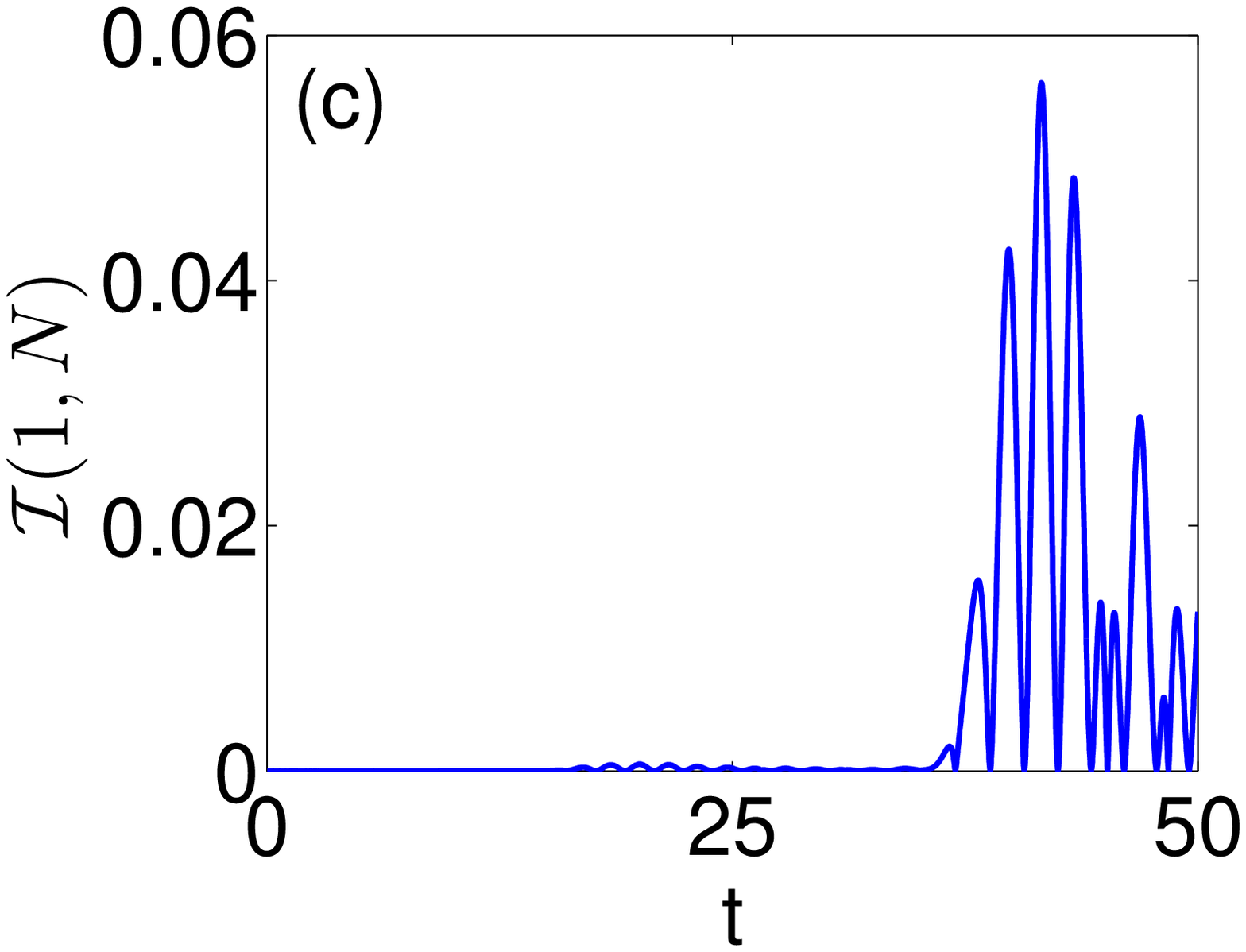}}
\subfigure{
    \includegraphics[width=4cm, clip]{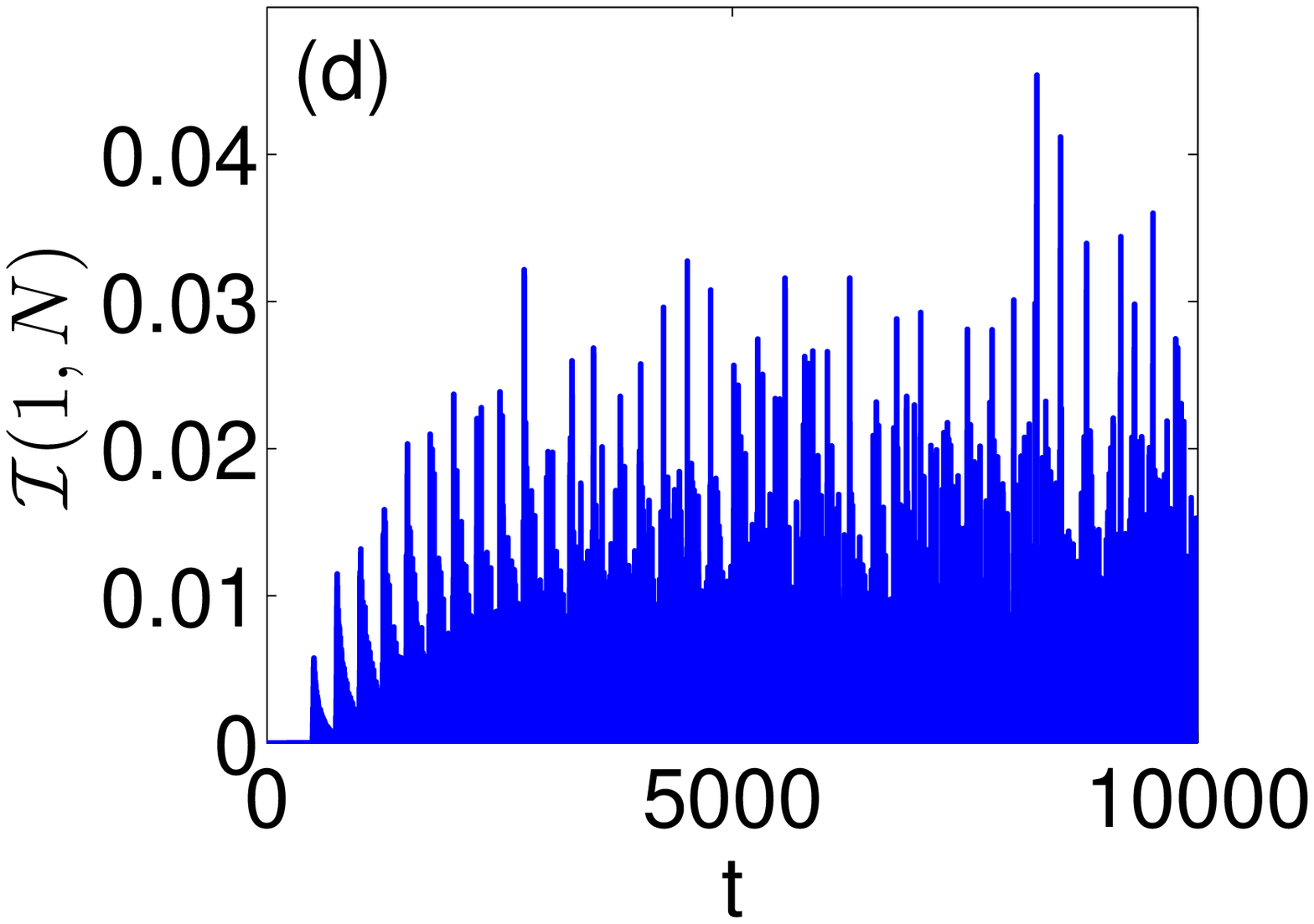}\label{fig:6d}}
\caption{Dynamics of MI of the spin chain with different lengths: (a) $N=20$,
(b) $N=30$, (c) $N=40$, (d) $N=500$.} \label{fig:6}
\end{figure}
\begin{figure}[tbh]
\centering
\subfigure{
    \includegraphics[width=4cm]{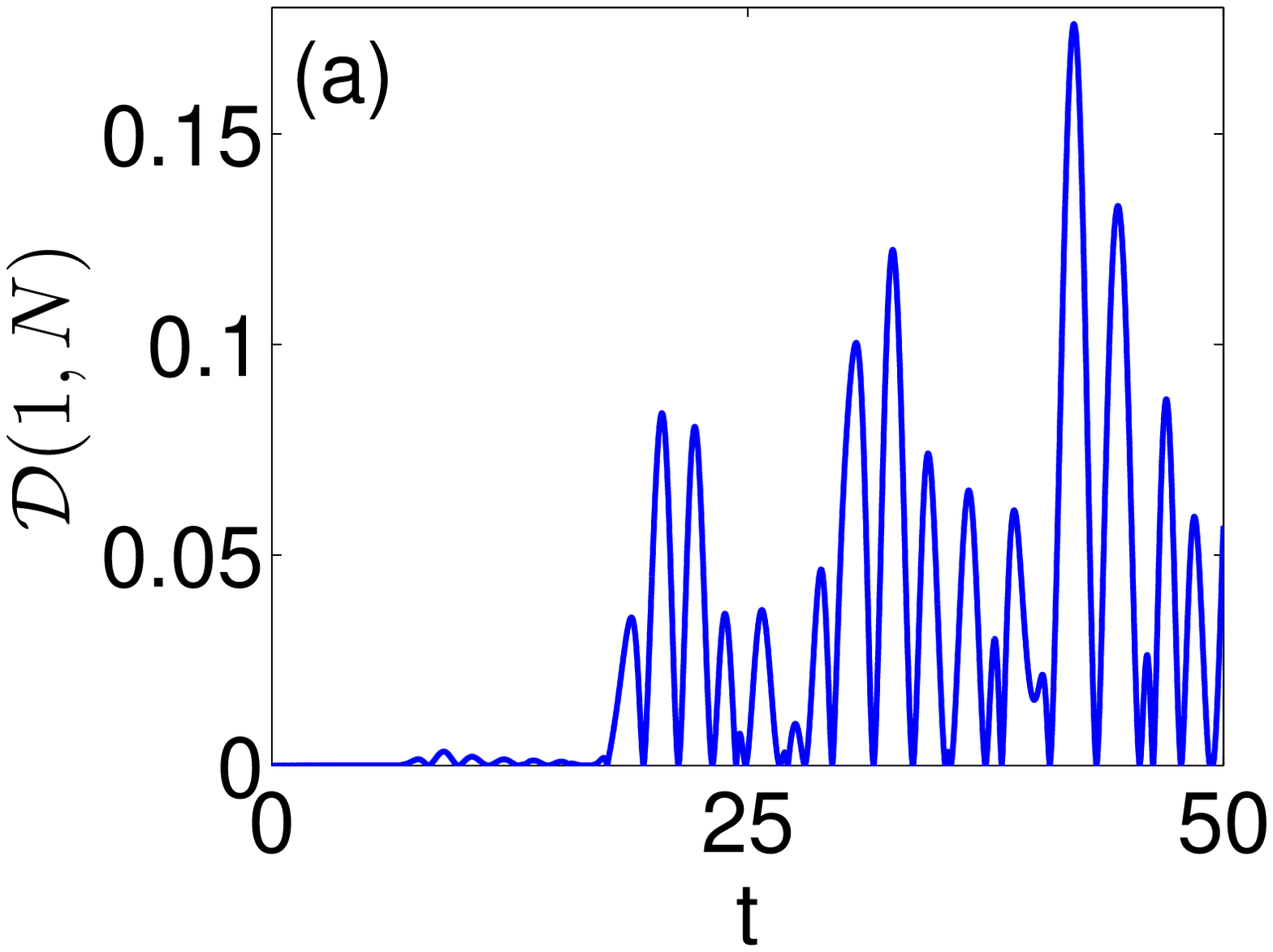}}
\subfigure{
    \includegraphics[width=4cm, clip]{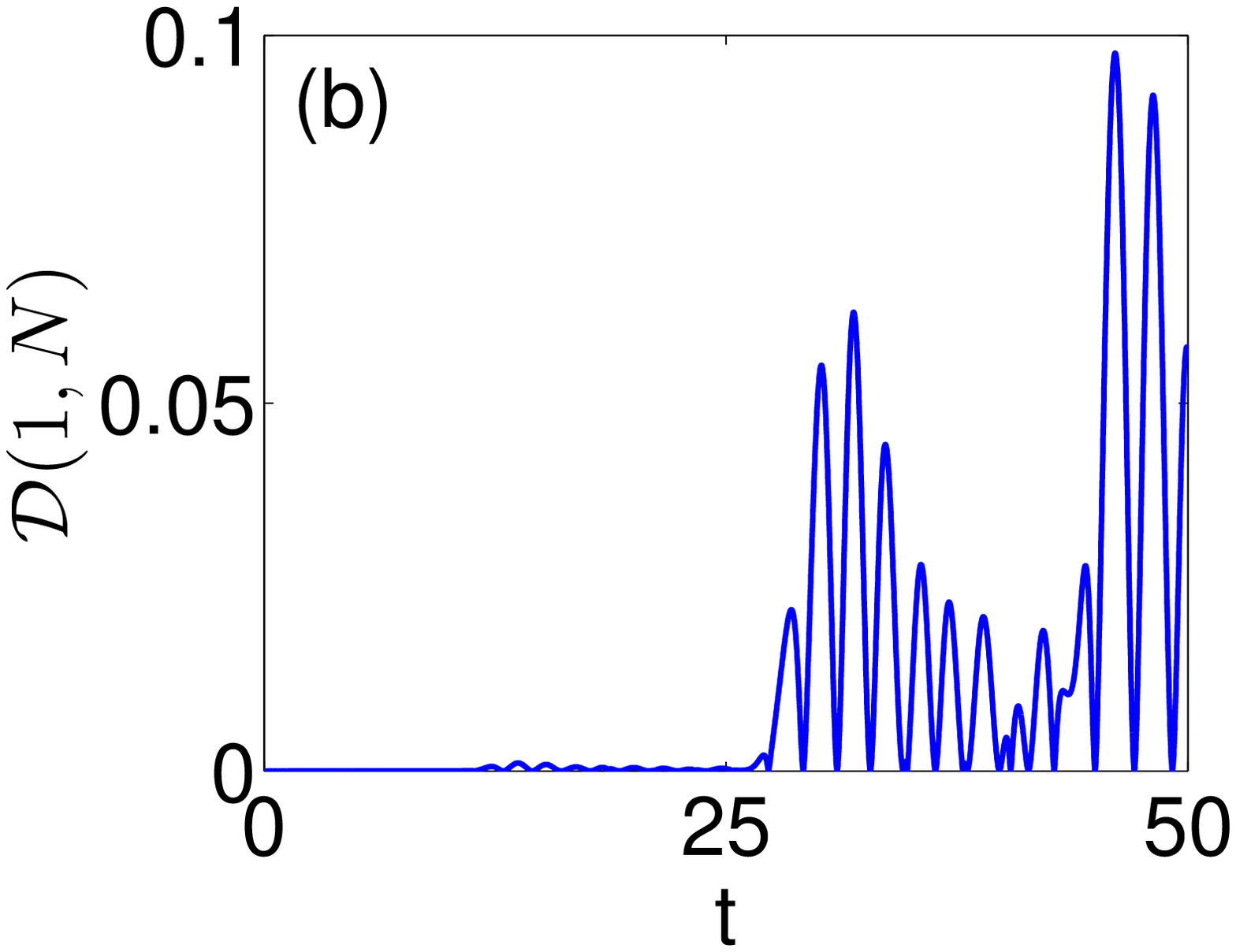}}
\subfigure{
    \includegraphics[width=4cm]{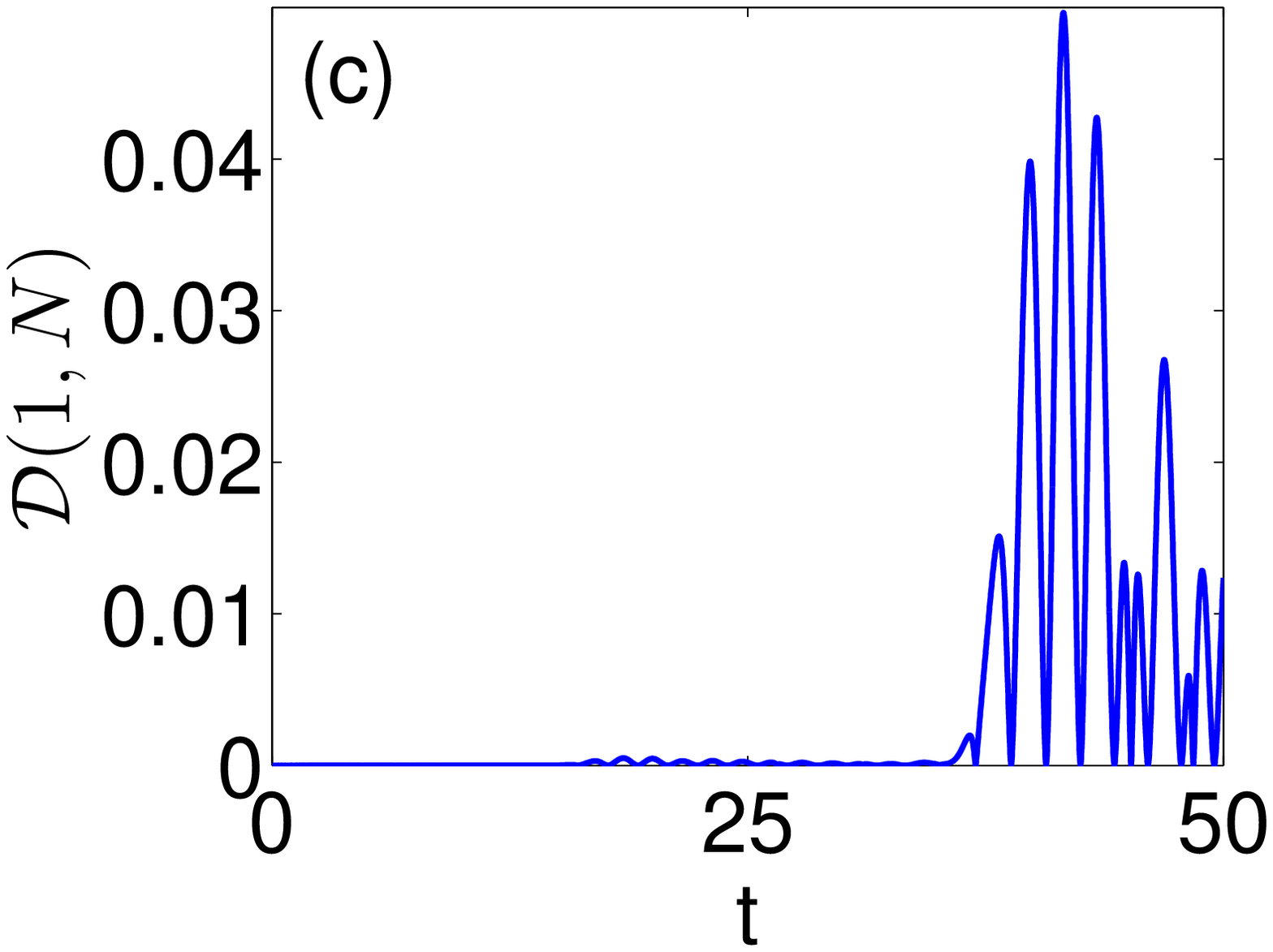}}
\subfigure{
    \includegraphics[width=4cm, clip]{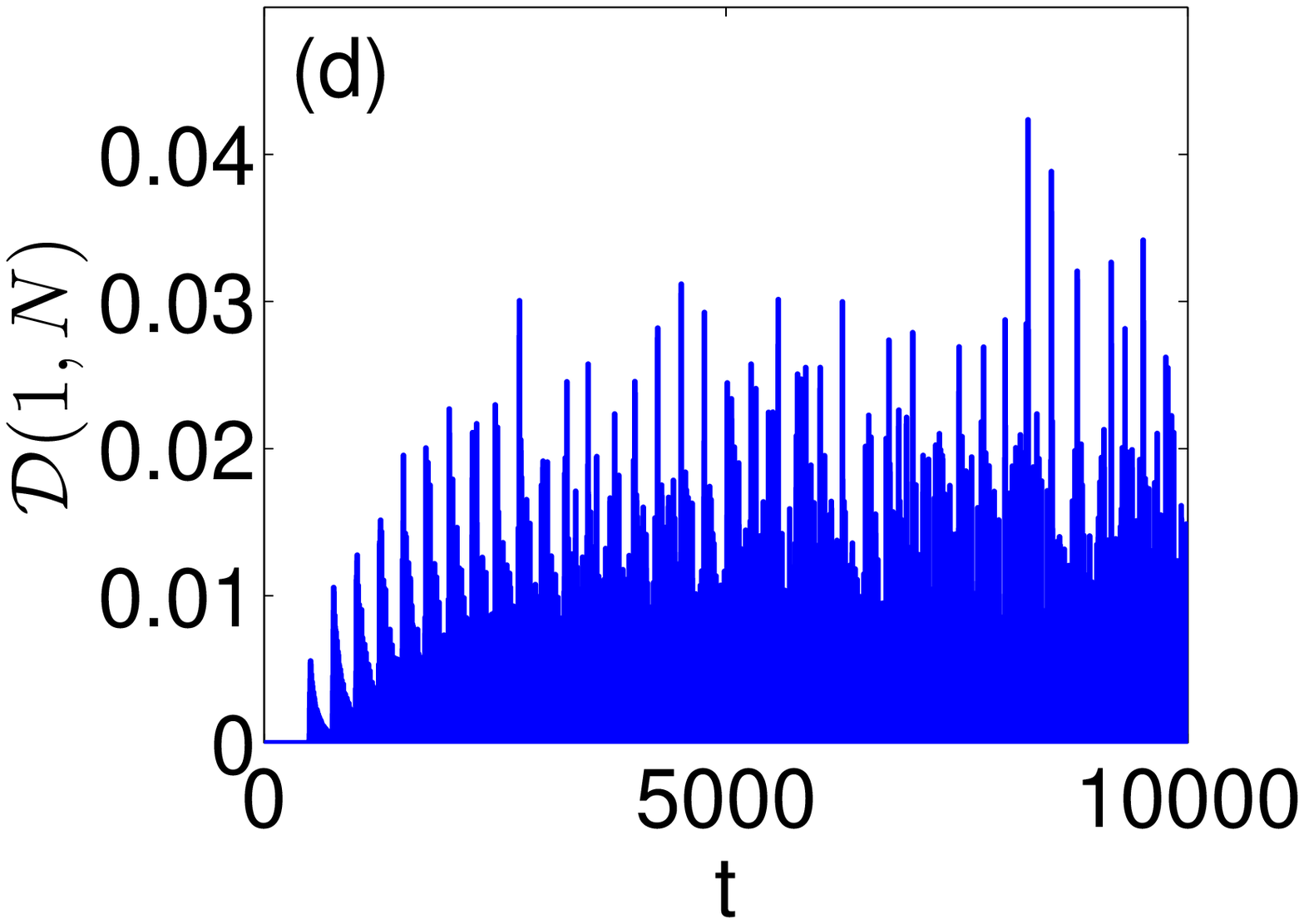}\label{fig:7d}}
\caption{Dynamics of QD of the spin chain with different lengths: (a) $N=20$,
(b) $N=30$, (c) $N=40$, (d) $N=500$.} \label{fig:7}
\end{figure}
\begin{figure}[tbh]
\centering
\subfigure{
    \includegraphics[width=4cm]{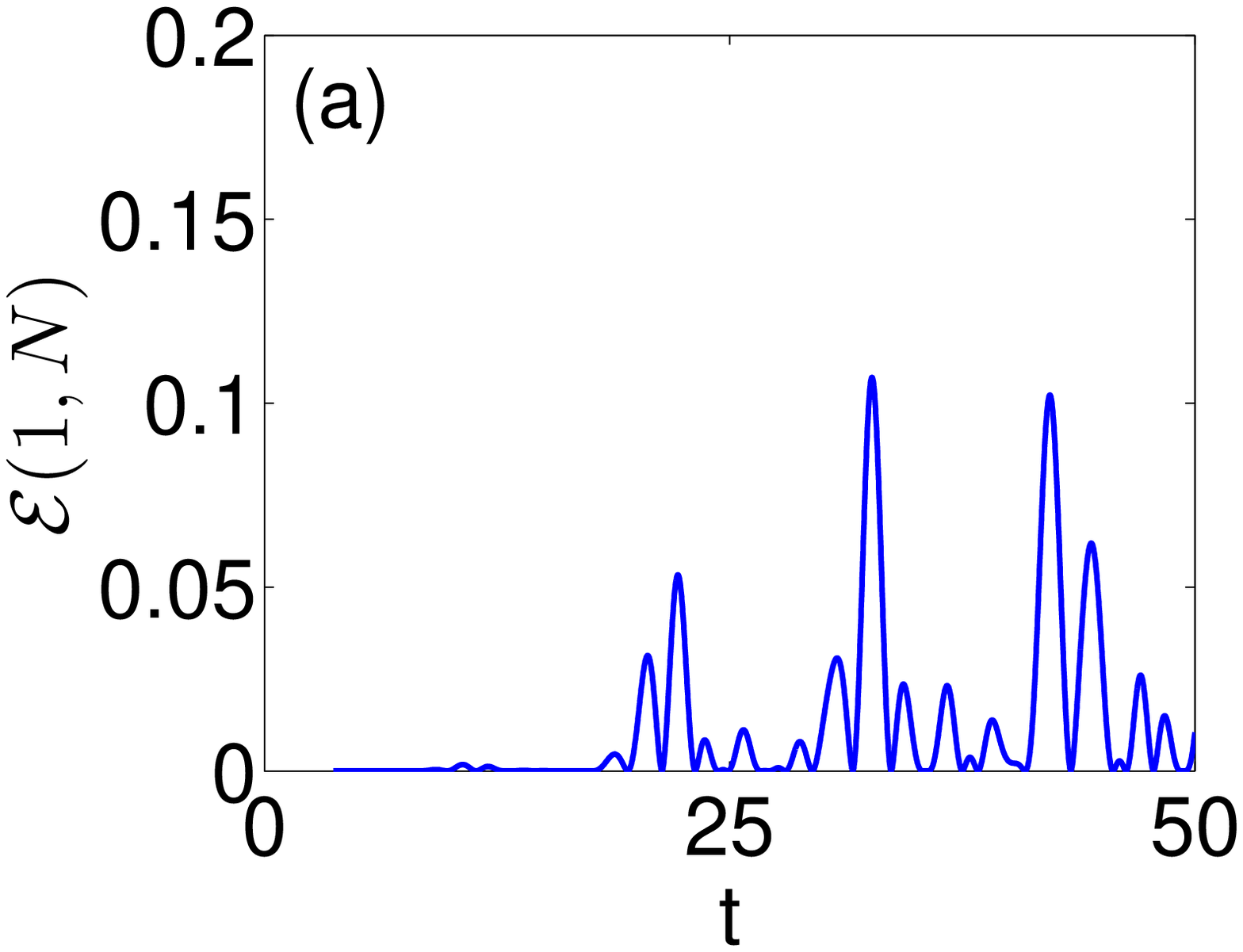}}
\subfigure{
    \includegraphics[width=4cm, clip]{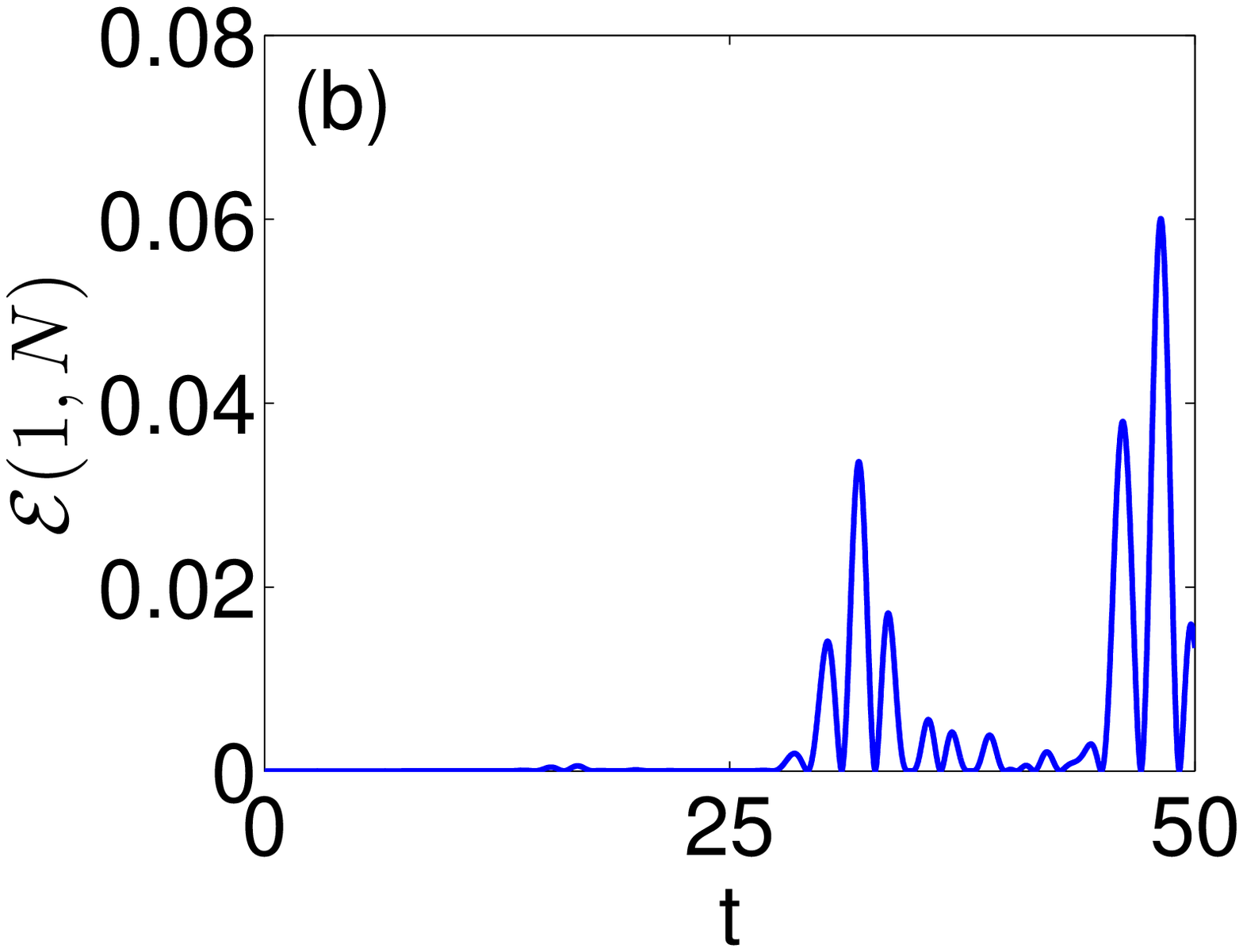}}
\subfigure{
    \includegraphics[width=4cm]{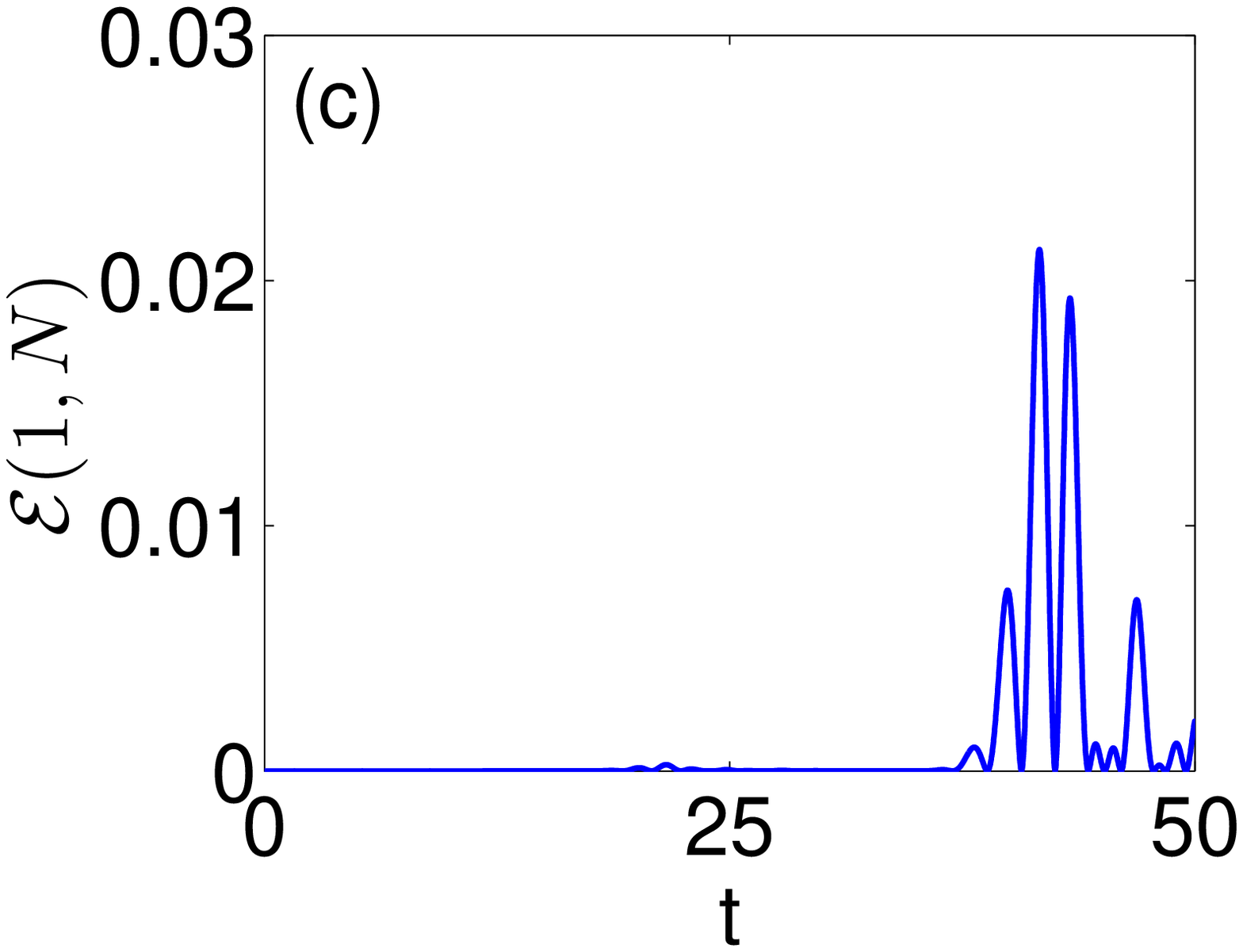}}
\subfigure{
    \includegraphics[width=4cm, clip]{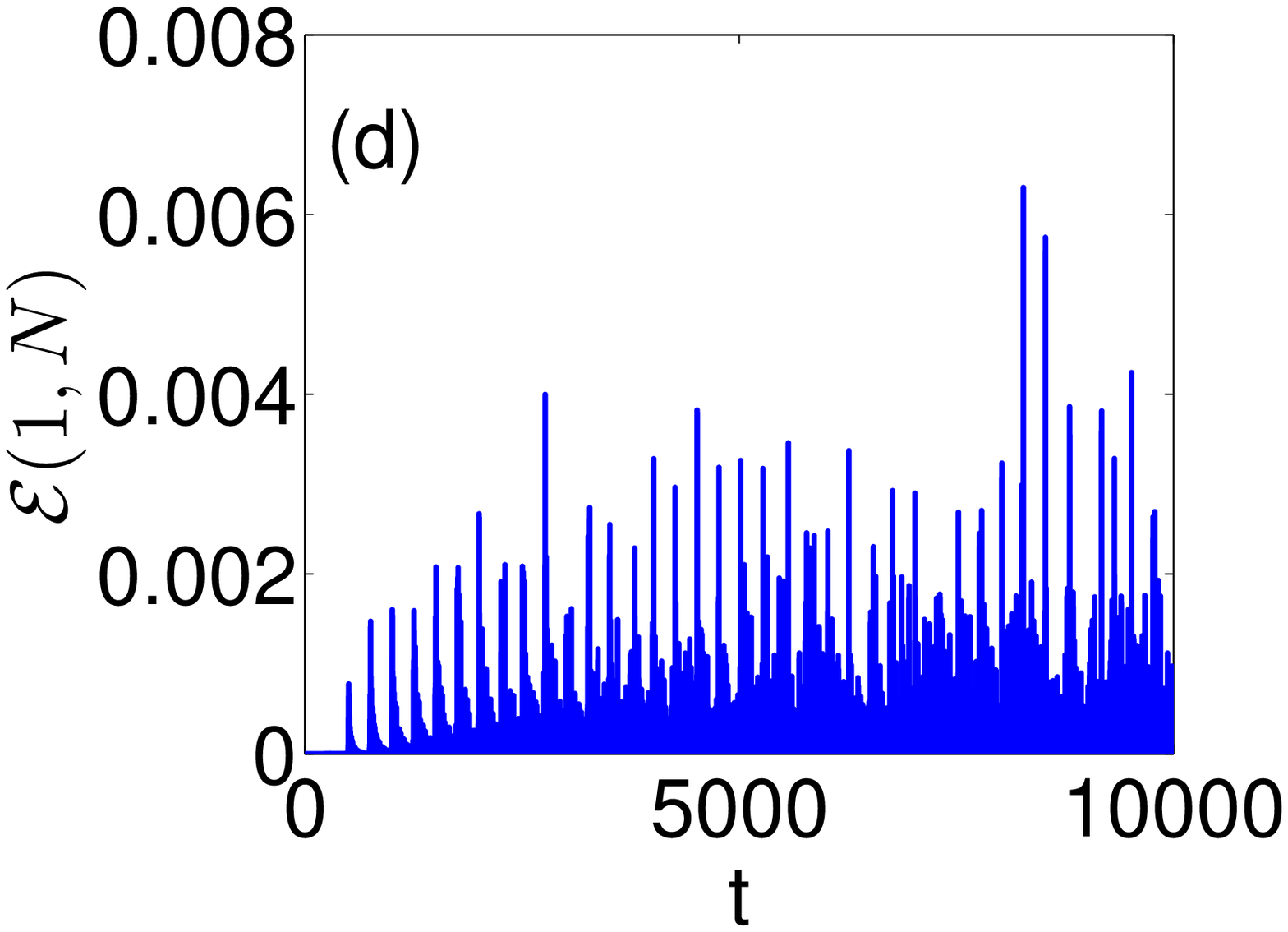}\label{fig:8d}}
\caption{Dynamics of EoF of the spin chain with different lengths: (a) $N=20$,
(b) $N=30$, (c) $N=40$, (d) $N=500$.} \label{fig:8}
\end{figure}

In Fig. \ref{fig:6}, Fig. \ref{fig:7}, and Fig. \ref{fig:8}, we
plot four cases in each figure, that is, $N=20$, $N=30$, $N=40$,
and $N=500$, in order to find the relation between the dynamical
evolution and the length of spin chain. From these three figures,
we observe that the dynamical evolutions of MI, QD and EoF are similar
with CFs, that is, the creation of MI, QD and EoF between the first
spin and the last spin is not instantaneous but requires the formation
time, and the longer the spin chain is, the more time to create correlations
takes. In addition, the maximum amplitudes that the MI, QD and EoF
can reach all become smaller with the increase of the length of spin
chain. However, the degrees of amplitude reduction are not same. For
example, we consider $N=500,$ from Fig.\ref{fig:6d}, Fig.\ref{fig:7d},
and Fig.\ref{fig:8d}, we can see that the maximum amplitudes of MI
and QD are about $4\times10^{-2}$, while the maximal amplitude of
EoF is only about $6\times10^{-3},$ which implies that MI and QD
are more robust to the length increase of the spin chain than EoF.

\section{Start-up time with different chain lengths\label{sec:4}}

In the above section we have shown that the time needed to create
correlations between the first spin and the last spin increases with
the increase of the spin chain length. However, we still do not know
how fast correlations can be created between the first spin and the
last spin in our concrete model, though there is an obvious and correct
answer: No faster than the Lieb-Robinson speed. In this section, we
will numerically study the relation between the start-up time of correlations
for different correlation measures and the length of spin chain. We
aim to find out the propagation speed of correlations and investigate
whether the propagation speed depends on the correlation measures.

\begin{figure}[t]
\centering
\subfigure{
    \includegraphics[width=4cm]{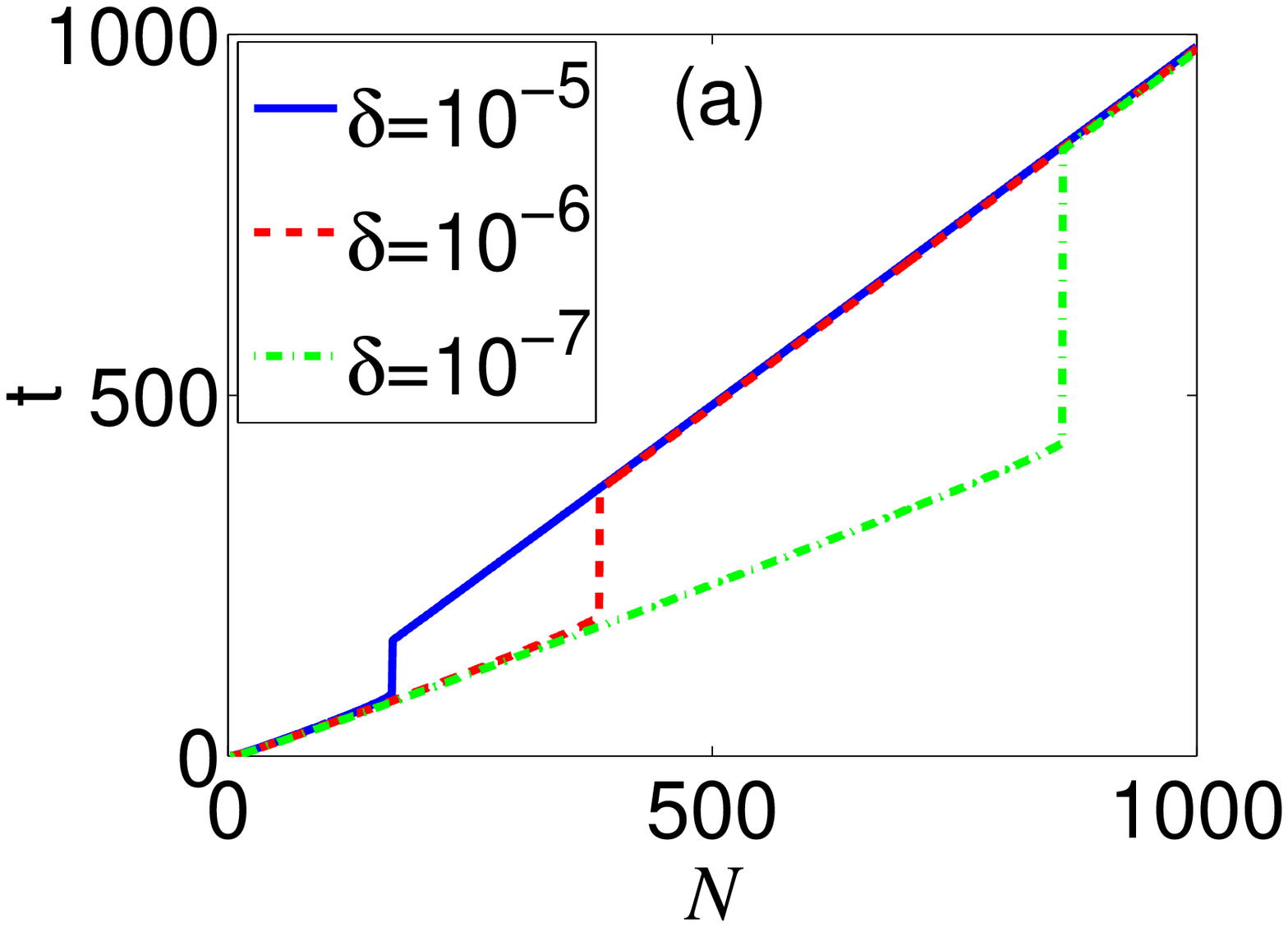}}
\subfigure{
    \includegraphics[width=4cm, clip]{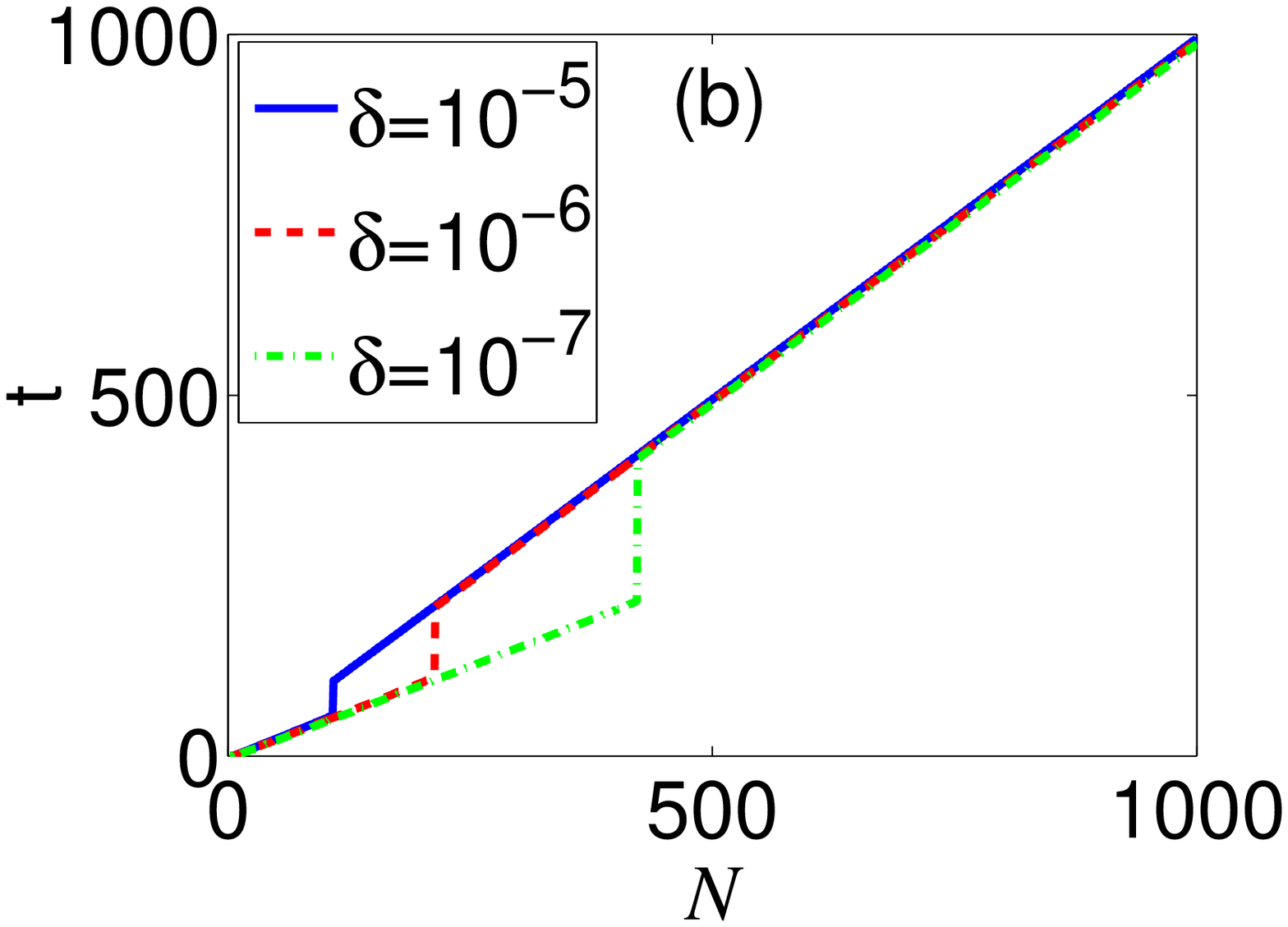}}
\subfigure{
    \includegraphics[width=4cm]{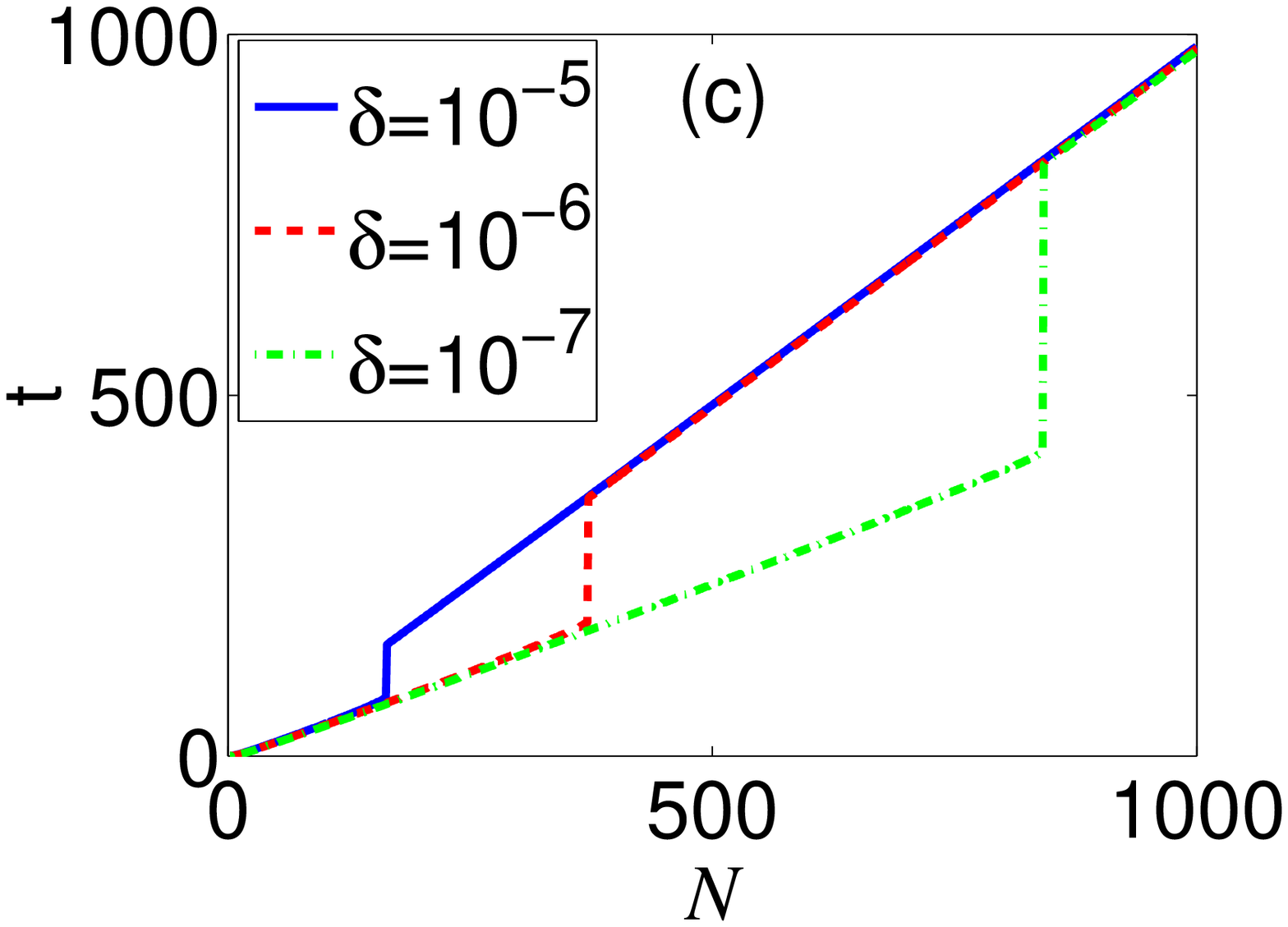}}
\subfigure{
    \includegraphics[width=4cm, clip]{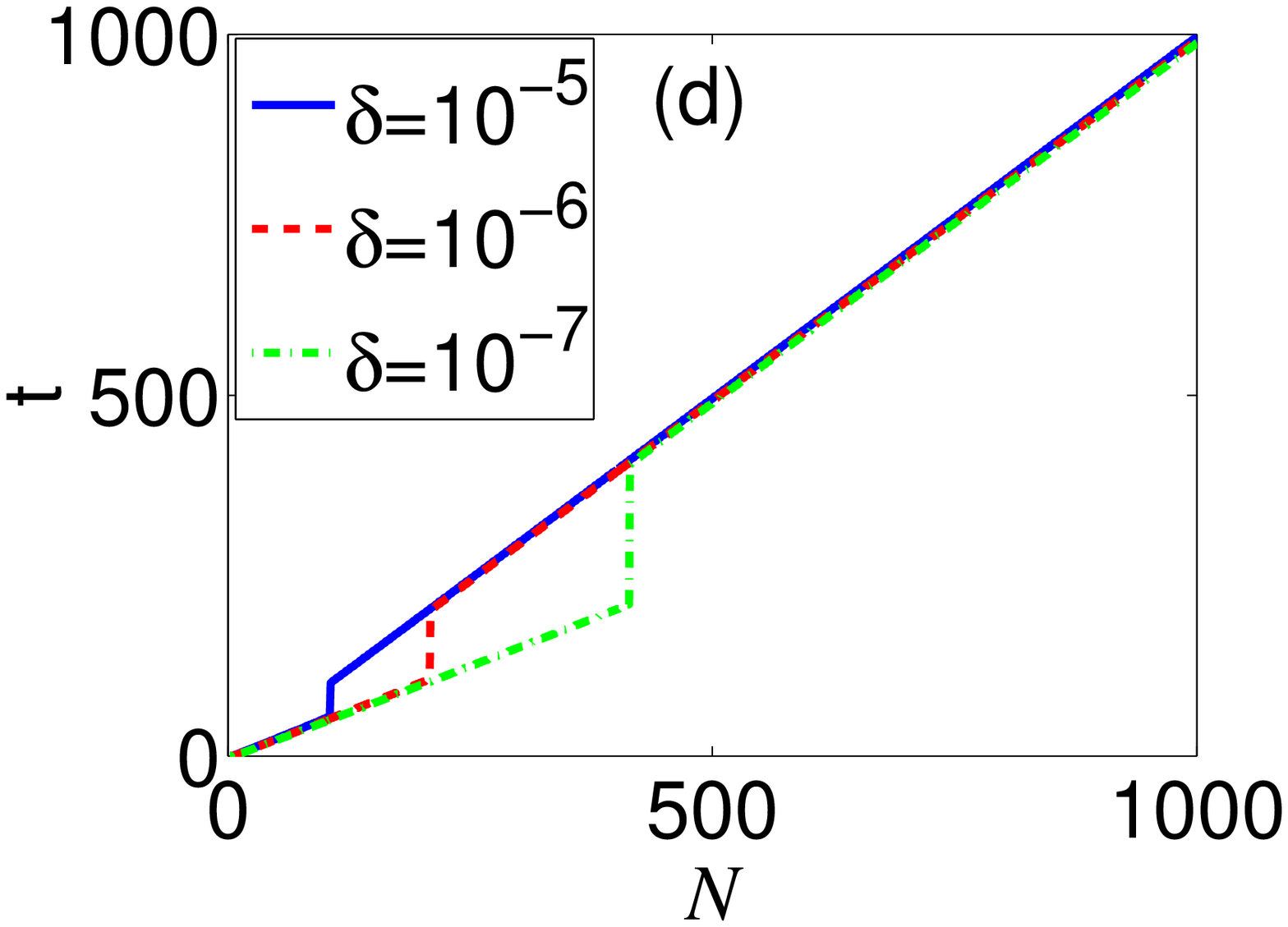}}
\caption{(Color online) Start-up time of (a) MI, (b) EoF, (c) QD, and (d) CC
as a function of $N$ with different criteria. The criteria of solid
line, dashed line, and dotted-dashed line are $\delta=10^{-4},$ $\delta=10^{-5},$
and $\delta=10^{-6}$, respectively.} \label{fig:9}
\end{figure}

\begin{figure}[h]
\centering
\subfigure{
    \includegraphics[width=6cm, clip]{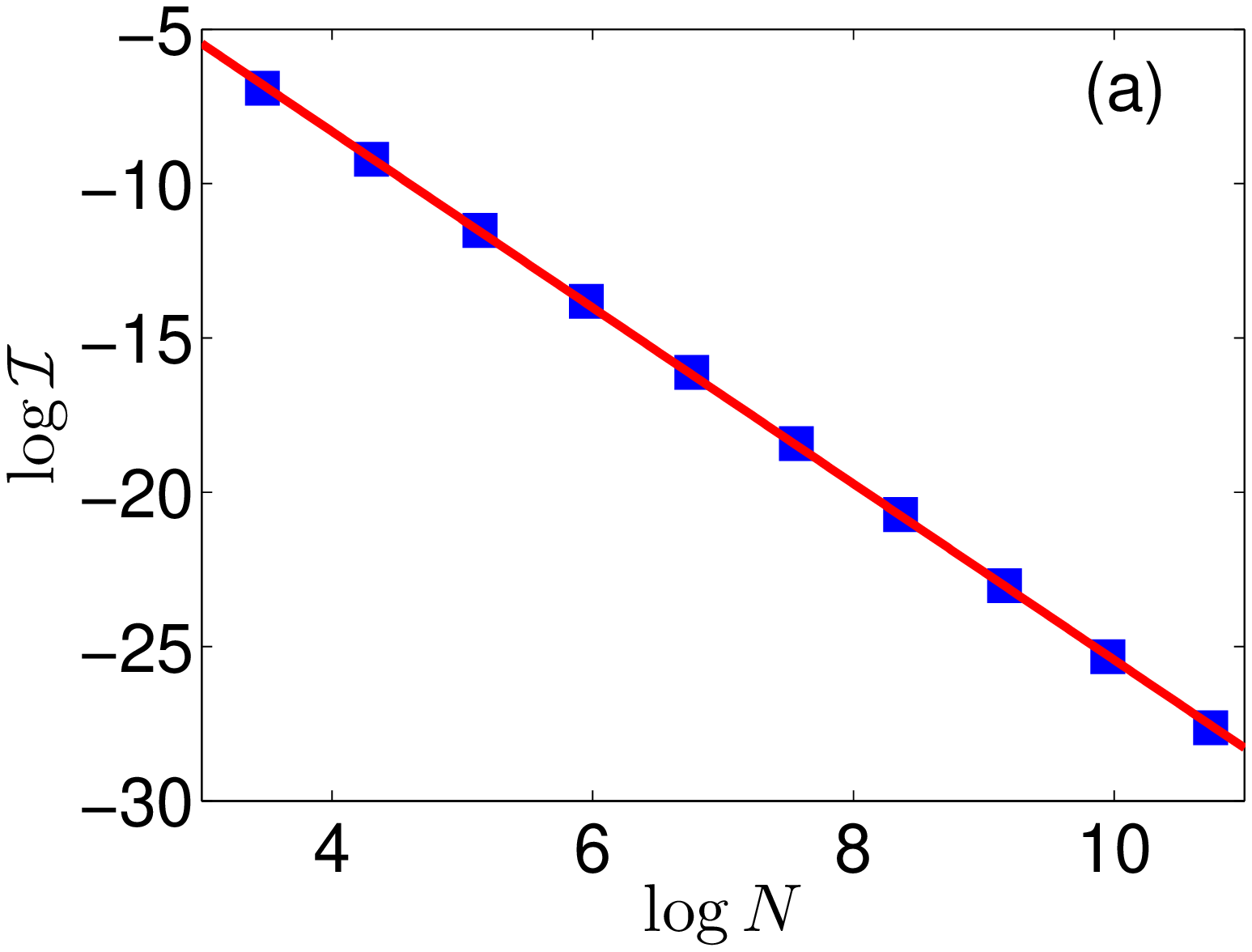}\label{fig:10a}}
\subfigure{
    \includegraphics[width=6cm, clip]{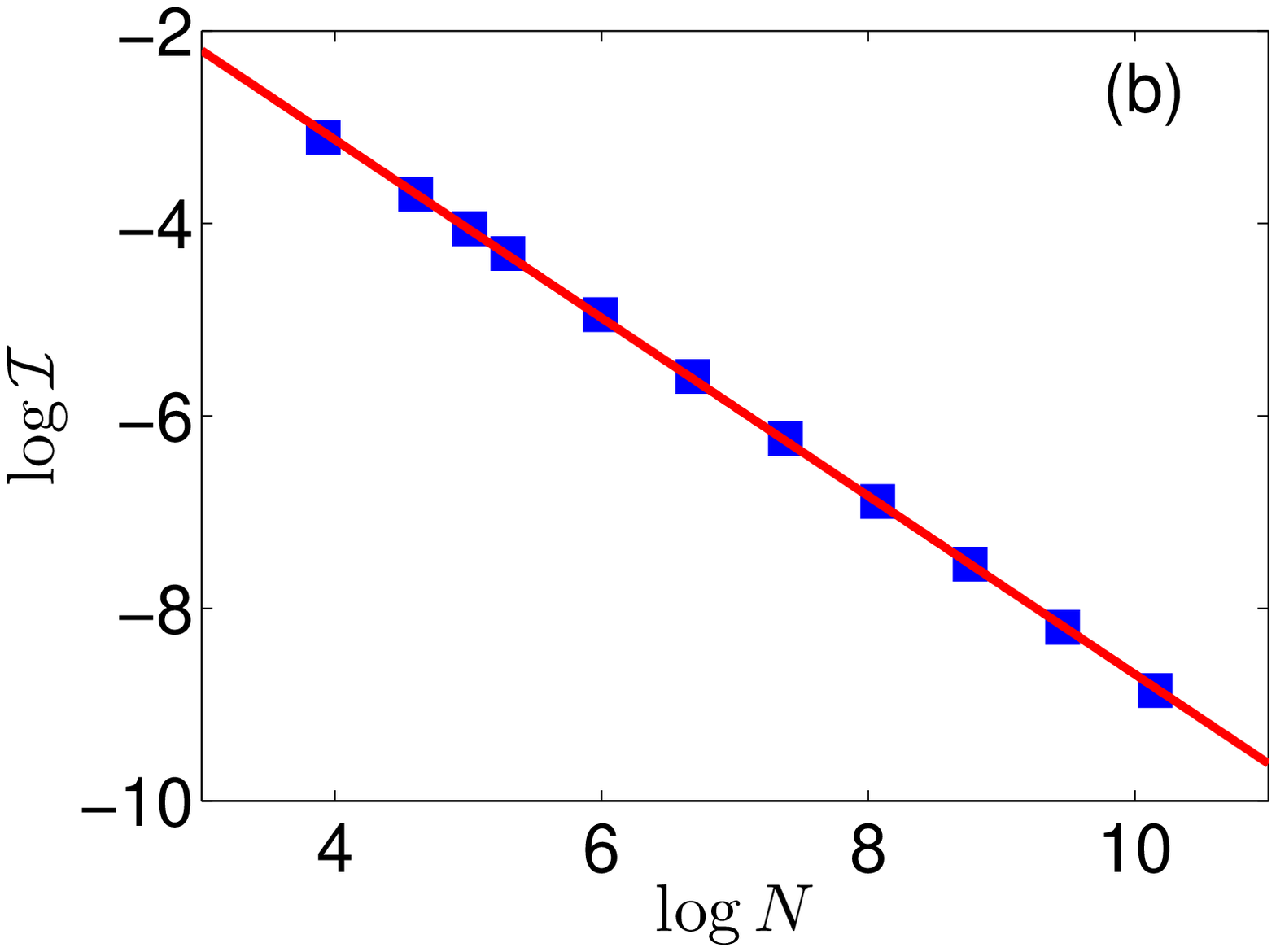}\label{fig:10b}}
\caption{$\log\mathcal{I}$ is linearly proportional to $\log N$ for (a) the
first peak value, (b) the second peak value.} \label{fig:10}
\end{figure}

Fig. \ref{fig:9} shows the start-up time of MI, QD, EoF, and CC as
a function of $N$, the length of spin chain, where $N$ is an integer
we choose from $2$ to $1000$. Here we define the criterion for correlation
appearance, which can be used to judge whether the correlations have
been created or not at a given time. In every subfigure of Fig. \ref{fig:9},
we plot three curves. Different curves have different criteria of
start-up $\delta$, that is, the criteria of solid line, dashed line,
and dotted-dashed line are $\delta=10^{-4},$ $\delta=10^{-5},$ and $\delta=10^{-6}$,
respectively. From Fig. \ref{fig:9}, we observe that the start-up
time as a function of the spin chain length has two segments, and
there is a sudden switch point, which moves rightwards with the decrease
of the criterion. From Fig. \ref{fig:6d}, Fig. \ref{fig:7d} ,and
Fig. \ref{fig:8d}, we can see that there are a series of obvious
peaks in the figures. In fact, the first segment in Fig. \ref{fig:9}
appears since the first peak reaches the criterion, and the second
segment appears since the second peak reaches the criterion. Thus
we plot the first peak value of MI as a function of the spin chain
length as shown in Fig. \ref{fig:10a}. We find the relation between
$\log\mathcal{I}$ and $\log N$ is linear, that is
\begin{equation}
\mathcal{I}=N^{\alpha}e^{\beta},
\end{equation}
where $\alpha=-2.8529$ and $\beta=3.0981$. We also plot the second
peak value of MI as a function of the spin chain length as shown in
Fig. \ref{fig:10b}, the relation between $\log\mathcal{I}$ and $\log N$
is also linear, but $\alpha=-0.9257$ and $\beta=0.5725$. So the
first peak will disappear sooner than the second peak with the increase
of $N$. This implies that the first peak becomes lower than the criterion
with the increase of $N$, but the second peak is still higher than
the criterion. This is the reason for the appearance of switch point
in Fig. \ref{fig:9}.

\begin{figure}[h]
\includegraphics[width=6 cm]{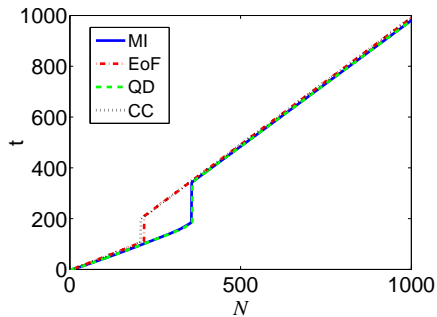}
\caption{(Color online) Comparison of start-up times of MI, QD, CC and EoF
with the criterion $\delta=10^{-6}$.} \label{fig:11}
\end{figure}

Lastly, in order to investigate whether the propagation speed depends
on the correlation measures, we plot them in the same figure with
the criterion $\delta=10^{-6}$ as shown in Fig. \ref{fig:11} to
compare the start-up times of MI, QD, CC and EoF. We find that the
propagation velocity does not depend on the correlation measures,
but the sudden switch point depends on the correlation measures.

In the above numerical results, the correlation start-up time is linearly
proportional to the length of the spin chain, which implies the existence
of the generalized Lieb-Robinson velocity. The sudden switch of the
generalized Lieb-Robinson velocities implies that different types
of correlations may have different generalized Lieb-Robinson velocities.
In the present model, the amplitudes of these two types of correlations
decreases exponentially with the length of the spin chain, and these
two types of correlations are characterized by different exponential
indexes. It is these different characteristics of correlations that
lead to the sudden switch of the Lieb-Robinson velocities.

\section{Discussion and Conclusion}

We have numerically investigated the problem on the generalized Lieb-Robinson
velocities in TFIC. We show that the generalized Lieb-Robinson velocity
can be observed by the dynamics of the correlations between the first
spin and the last spin. A sudden switch of the generalized Lieb-Robinson
velocities appears in TFIC. Here a main open problem is whether the
phenomena in TFIC are model independent. However, the method we adopt
for TFIC uses the analytical results, which is not available for a
general spin model. In addition, as far as we know, the algorithms
for quantum dynamics are not powerful enough to simulating for a sufficient
long time with a long spin chain. In this direction, we study another
related mode, the isotropic Heisenberg chain.

\begin{figure}[h]
\centering
\subfigure{
    \includegraphics[width=4cm,clip]{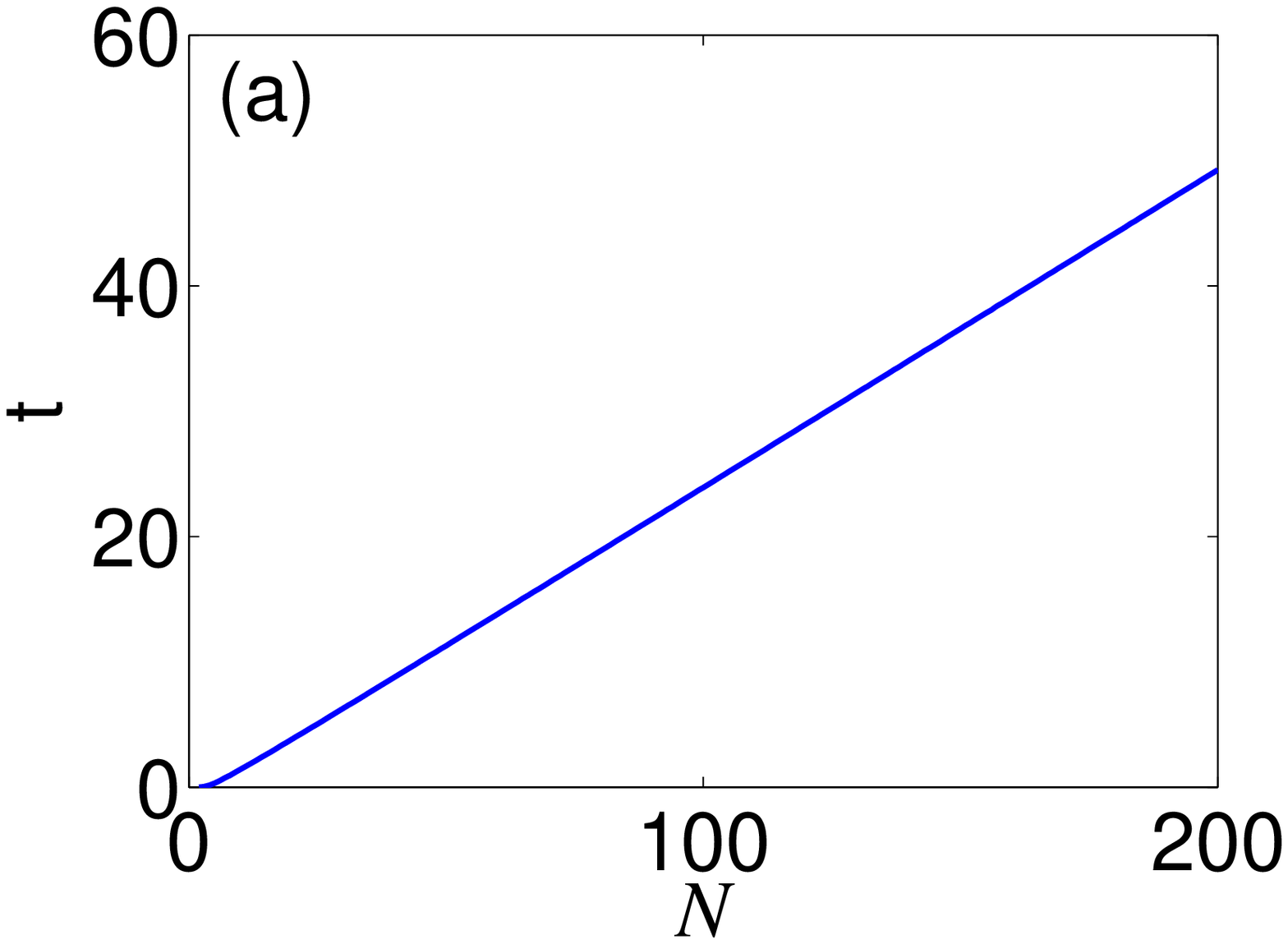}}
\subfigure{
    \includegraphics[width=4.2cm, clip]{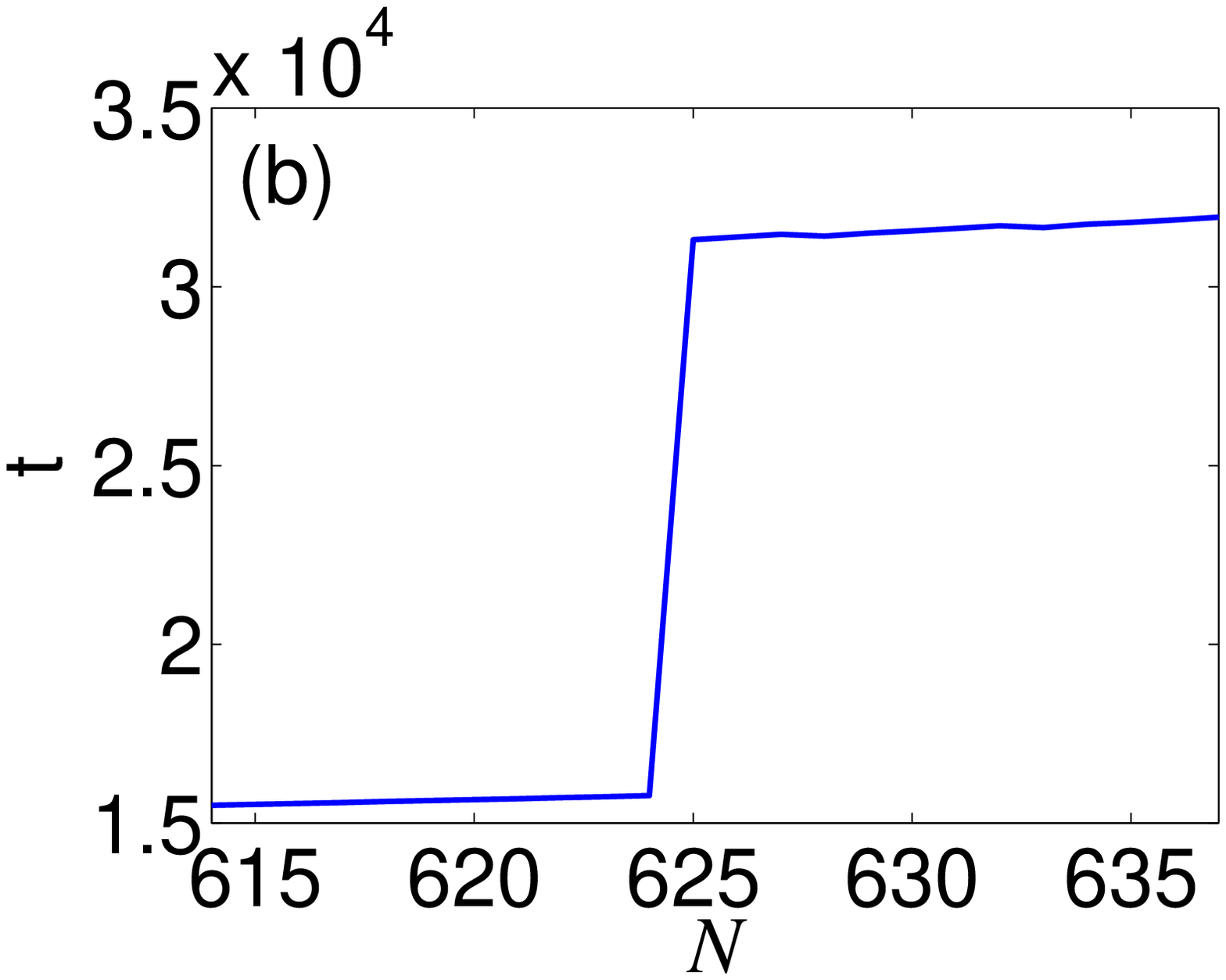}}
\caption{Start-up time of CF of the Heisenberg chain with uniform coupling
as a function of $N$: (a) from $N=2$ to $N=200$, (b) from $N=614$
to $N=637$.}\label{fig:12}
\end{figure}

The start-up time of CF of isotropic Heisenberg chain with uniform
coupling as a function of the chain length $N$ is demonstrated in
Fig. \ref{fig:12}. We find that the phenomena on the generalized
Lieb-Robinson velocity have the same characteristic in the isotropic
Heisenberg chain.

In conclusion, we have numerically investigated the generalized Lieb-Robinson
velocity in an Ising spin chain with a strong transverse field through
studying the dynamical evolution of correlations. The generalized
Lieb-Robinson velocities are demonstrated in different types of correlation
measures, which include correlation function, mutual information,
quantum discord, and entanglement of formation. We find that one of
the correlation functions shows a special behavior depending on the
parity of the spin number. In particular, we find that there is a
switch in the generalized Leb-Robinson velocities, which implies that
different types of correlations may show different generalized Lieb-Robinson
velocities.

\begin{acknowledgments}
This work is supported by NSF of China under Grants No. 10975181,
No. 11175247, No. 11105020, and No. 11005013. Yu Guo is also supported
by  the China Postdoctoral Science Foundation funded project.
\end{acknowledgments}


\end{document}